\documentclass[twocolumn]{aastex701}

\newcommand{\kms}
{\ensuremath{\mathrm{km\,s^{-1}}}}

\begin{document}

\title{NOEMA Observations of the Center of M33: A Parsec-scale Molecular Filament and an Ordered Gas--Star-Formation Sequence}

\author[0000-0002-7172-6306]{Zongnan Li}
\email{zongnanli@kasi.re.kr}
\affiliation{Korea Astronomy and Space Science Institute, 776 Daedeok-daero, Yuseong-gu, Daejeon 34055, Republic of Korea}
\affiliation{National Astronomical Observatory of Japan, 2-21-1 Osawa, Mitaka, Tokyo, 181-8588, Japan}
\affiliation{East Asian Core Observatories Association (EACOA) Fellow}
\author[0000-0003-0355-6437]{Zhiyuan Li}
\email{lizy@nju.edu.cn}
\affiliation{School of Astronomy and Space Science, Nanjing University, Nanjing 210023, China}
\affiliation{Key Laboratory of Modern Astronomy and Astrophysics, Nanjing University, Nanjing 210023, China}
\author[0000-0002-1583-8514]{Gan Luo}
\email{}
\affiliation{Institut de Radioastronomie Millim\'etrique, Grenoble, France}
\author[0000-0003-4603-7119]{Sheng-Yuan Liu}
\email{}
\affiliation{Institute of Astronomy and Astrophysics, Academia Sinica, 11F of ASMAB, AS/NTU No.1, Sec. 4, Roosevelt Road, Taipei 10617, Taiwan}
\author{Sihan Jiao}
\email{}
\affiliation{National Astronomical Observatories, Chinese Academy of Sciences, A20 Datun Road, Chaoyang District, Beijing 100101, China}
\affiliation{Max Planck Institute for Astronomy, K\"onigstuhl 17, D-69117 Heidelberg, Germany}
\author[0000-0002-5927-2049]{Fanyi Meng}
\email{}
\affiliation{New Cornerstone Science Laboratory, Department of Astronomy, Tsinghua University, Beijing, 100084, China}
\affiliation{School of Astronomy and Space Science, University of Chinese Academy of Sciences, Beijing 100049, China}

\author{Zhiyuan Ren}
\email{}
\affiliation{National Astronomical Observatories, Chinese Academy of Sciences, A20 Datun Road, Chaoyang District, Beijing 100101, China}
\author{Daisuke Iono}
\email{d.iono@nao.ac.jp}
\affiliation{National Astronomical Observatory of Japan, 2-21-1 Osawa, Mitaka, Tokyo 181-8588, Japan}
\author{Jiachang Zhang}
\email{zhangjiachang@smail.nju.edu.cn}
\affiliation{School of Astronomy and Space Science, Nanjing University, Nanjing 210023, China}
\affiliation{Key Laboratory of Modern Astronomy and Astrophysics, Nanjing University, Nanjing 210023, China}
\author[0000-0003-3010-7661]{Di Li}
\email{}
\affiliation{New Cornerstone Science Laboratory, Department of Astronomy, Tsinghua University, Beijing, 100084, China}
\affiliation{National Astronomical Observatories, Chinese Academy of Sciences, A20 Datun Road, Chaoyang District, Beijing 100101, China}
\author[0000-0002-9390-9672]{Chao-Wei Tsai}
\email{}
\affiliation{National Astronomical Observatories, Chinese Academy of Sciences, A20 Datun Road, Chaoyang District, Beijing 100101, China}
\author{Jongsoo Kim}
\email{}
\affiliation{Korea Astronomy and Space Science Institute, 776 Daedeok-daero, Yuseong-gu, Daejeon 34055, Republic of Korea}

\begin{abstract}

We present NOEMA CO(1--0) observations of the inner $\sim$500 pc of M33 at a resolution of $0.91''\times0.37''$ ($3.7\times1.5$ pc), resolving the compact molecular gas structures and revealing their relation to recent massive star formation. The most prominent structure is a southeastern molecular filament (SE filament) that extends 76 pc along its principal axis, is 23 pc wide, and contains three CO-bright knots. Across eight independent transverse cuts, the adjacent H$\alpha$ ridge is displaced preferentially to one side of the principal CO ridge by a median \(10.0^{+0.9}_{-1.4}\) pc. Young OB candidates are more closely associated with H$\alpha$ than with compact CO. The filament shows little velocity change along its length but a resolved gradient across its minor axis.
Its elongated morphology, transverse kinematics, and ordered CO--H$\alpha$ distribution are consistent with locally asymmetric molecular-gas compression and subsequent massive star formation. Across the wider field, 23 well-measured compact clumps have a median radius of 2.67 pc and a median velocity dispersion of 2.15 \kms. Their linewidth and apparent virial-parameter distributions show no clear difference between the SE filament and the rest of the field. Although the local spiral potential and pattern speed remain unconstrained, the observations provide a resolved view of a compression-compatible molecular-to-ionized-gas sequence in the center of M33.

\end{abstract}


\section{Introduction}

Molecular clouds provide the immediate material for star formation, a key process that shapes galaxies and regulates their evolution through subsequent stellar feedback. Their evolution is governed by self-gravity, pressure, turbulence, and magnetic fields \citep{2007ARA&A..45..565M,2020SSRv..216...50C} and is modulated by galactic environment and large-scale dynamics {\citep{2019ApJ...883....2S,2021MNRAS.505.4048L,2026ApJS..282...39J}}. Molecular-cloud structure can therefore reveal how galactic-scale gas flows affect the conditions for star formation on parsec and subparsec scales. A substantial fraction of molecular gas is organized into filaments, and star-forming clumps and cores are often concentrated along these structures. Giant molecular filaments (GMFs), with lengths of tens to hundreds of parsecs and masses of $10^4$--$10^5~M_\odot$, may therefore preserve information about both molecular-cloud assembly and the onset of clustered star formation {\citep{2014A&A...568A..73R,2016ApJS..226....9W,2018ApJ...864..153Z}}. How such long, coherent molecular structures form within galactic disks remains uncertain.

On galactic scales, elongated molecular structures may be assembled by compression associated with spiral structure. In the classical quasi-stationary density-wave picture, a long-lived spiral pattern rotates with a nearly fixed pattern speed while gas flows through the spiral potential and is compressed in a shock \citep{1964ApJ...140..646L,1969ApJ...158..123R}. Molecular gas should then accumulate near the compressed ridge, with massive star formation appearing downstream after a delay. Dynamic spiral models instead describe arms as transient, recurrent structures that approximately co-rotate with the disk. Gas can converge toward such an arm from both sides, so a persistent one-directional molecular-gas-to-star-formation sequence is not generally expected \citep{2014PASA...31...35D,2016MNRAS.460.2472B}. {Recent arm--interarm comparisons suggest a complementary ``gatherer'' picture, in which spiral arms concentrate molecular gas and star-forming regions without necessarily altering cloud lifetimes or strongly enhancing star-formation efficiency \citep{2024ApJ...973..137S,2025A&A...698A.296R}.}
The relative distributions of molecular gas and recent star formation provide an observational constraint on these flow geometries. Ordered CO--H$\alpha$ offsets have been reported across the arms of the grand-design galaxy M51 and in other nearby spiral galaxies \citep{2009ApJ...697.1870E,2013ApJ...763...94L}, consistent in some cases with gas compression followed by massive star formation. Such offsets are not detected uniformly across spiral galaxies or even along individual spiral arms. Their appearance depends on the coherence of the arm, the tracers, spatial resolution, and the gas motion relative to the spiral pattern. More generally, CO-bright, overlapping CO--H$\alpha$, and H$\alpha$-bright regions statistically represent different stages of molecular-cloud evolution, whose durations depend on the local galactic environment \citep{2022ApJ...927....9P}. A local offset is therefore most physically informative when it maintains a consistent direction along an extended structure and is considered together with the gas morphology, kinematics, and galactic environment.

M33 therefore provides a useful laboratory for testing whether such an ordered sequence can arise without a prominent grand-design spiral pattern. It is the nearest actively star-forming late-type spiral galaxy outside the Milky Way: at a distance of approximately 840 kpc, \(1\arcsec\) corresponds to about 4 pc \citep{2001ApJ...553...47F}. Its moderate inclination of \(55^\circ\) also provides a comparatively unobscured external view at physical scales comparable to those accessible in Galactic studies \citep{2018MNRAS.479.2505K}. M33 has a flocculent spiral morphology, and simulations show that gravitational instabilities, gas physics, and stellar feedback can reproduce much of its structure without a long-lived global density wave \citep{2018MNRAS.478.3793D}. Parsec-scale observations of M33 therefore allow us to examine whether locally ordered molecular gas and recent star formation can arise within its dynamically evolving disk.

The central region of M33 also provides an environment distinct from both its outer disk and the Central Molecular Zone (CMZ) of the Milky Way. The Milky Way CMZ contains warm, turbulent, high-pressure molecular gas with velocity dispersions substantially above those of disk clouds \citep{1996ARA&A..34..645M,2017A&A...603A..89K}. M33, in contrast, is a low-mass, bulgeless galaxy with no detected supermassive black hole \citep{2001AJ....122.2469G}. 
Its parsec-scale molecular structures therefore offer a way to examine gas dynamics close to a galactic center without the deep nuclear potential of the Milky Way CMZ.

Previous CO observations have established the molecular-gas distribution and cloud population of M33 over a wide range of scales. The BIMA CO(1--0) survey showed that GMCs are closely associated with H{\sc i} filaments and H{\sc ii} regions \citep{2003ApJS..149..343E}. Nobeyama and combined Nobeyama--BIMA--FCRAO observations mapped the CO(1--0) distribution and its radial variations \citep{2007ApJ...654..240R,2011PASJ...63.1171T,2012PASJ...64..133O}. The IRAM 30 m CO(2--1) survey examined the association between GMCs and young stellar populations \citep{2014A&A...567A.118D,2017A&A...601A.146C,2018A&A...612A..51B}, while the ACA+IRAM survey cataloged 848 molecular clouds at approximately 30 pc resolution \citep{2023ApJ...953..164M,2024PASJ...76.1098K}. These studies found that more massive and more gravitationally bound clouds are preferentially associated with active high-mass star formation, and interpreted systematic changes in cloud properties with H{\sc ii}-region association as an evolutionary sequence.\ Collectively, these surveys characterize GMC-scale populations but do not resolve their parsec-scale internal structure.

A particularly relevant parsec-scale comparison is provided by ALMA observations of GMC-16 in the M33 disk. At $1$--$2$ pc resolution, these observations revealed several $50$--$70$ pc molecular filaments and a $10$--$20$ pc displacement between the filaments and downstream H{\sc ii} regions. Their morphology, position relative to the direction of disk rotation, and gas kinematics were interpreted as evidence for formation by a galactic spiral shock \citep{2020ApJ...896...36T}. 
No comparable parsec-scale CO study has yet focused on the central region of M33.

In this work, we present NOEMA CO(1--0) observations of the inner \(\sim500\) pc of M33 at a spatial resolution of a few parsecs. We first describe the compact molecular component detected across the full field and then focus on a prominent southeastern molecular filament. We measure its morphology, kinematics, compact-clump population, and spatial relationship with H$\alpha$ emission and young OB-star candidates. Our main goal is to determine whether the ordered molecular-gas and recent-star-formation distribution is consistent with molecular-gas compression followed by massive star formation. As a complementary analysis, we compare compact clumps within and outside the SE filament to assess whether the SE environment is accompanied by measurable clump-scale differences. We retain dynamic spiral structure, cloud--cloud interaction, and stellar feedback as alternative interpretations.
The paper is organized as follows. The NOEMA observations, data reduction, and ancillary data are described in Section~\ref{sec:obs}. {Section~\ref{sec:result} presents the whole-field CO distribution, the compact-clump decomposition, and the morphology and kinematics of the SE filament.} In Section~\ref{sec:discuss}, we discuss the possible origin of the filament and place the compact clumps in their broader environmental context. Our main findings are summarized in Section~\ref{sec:sum}.

\section{Observations and Data Reduction}\label{sec:obs}

\subsection{NOEMA CO(1--0) observations and data reduction}

\begin{figure*}[htpb!]
    \centering
    \includegraphics[width=\linewidth]{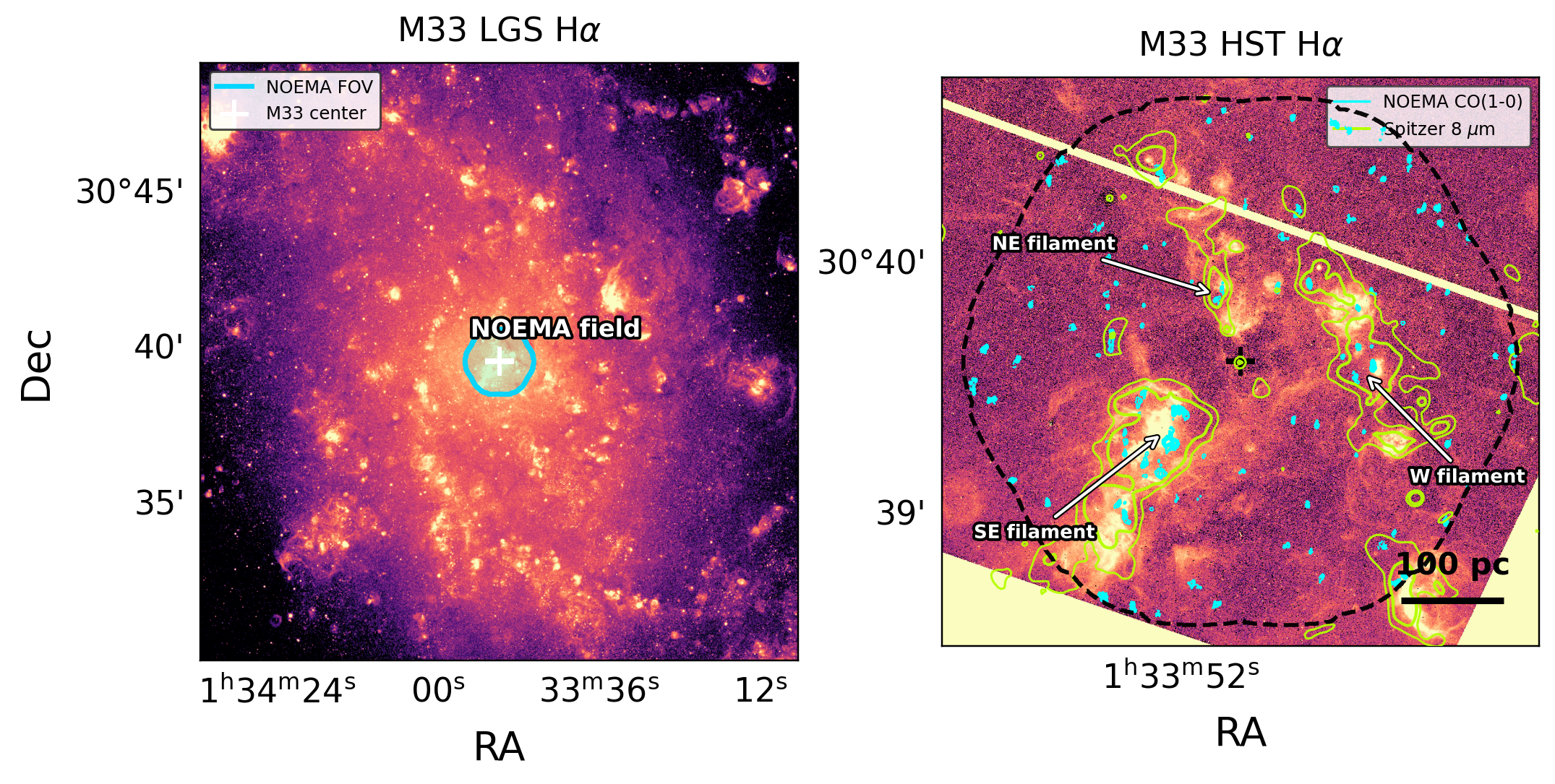}
    \caption{Global context and local morphology of the central NOEMA
    field. Left: Local Group Galaxy Survey H$\alpha$ image of M33
    \citep{2006AJ....131.2478M,2007AJ....134.2474M}. The cyan outline marks the NOEMA field, and the cross marks the adopted M33 center.
    Right: continuum-subtracted HST H$\alpha$ image of the NOEMA field. Cyan contours show NOEMA CO(1--0) integrated intensity with levels \(0.04,\ 0.07,\ 0.12,\ 0.20,\ 0.32\ {\rm Jy\,beam^{-1}\,km\,s^{-1}}\), yellow contours show Spitzer/IRAC 8~$\mu$m emission at \(7.4\) and \(11.1~{\rm MJy\,sr^{-1}}\) \citep{2009ApJ...703..517D}, and the black dashed curve marks the NOEMA footprint. The arrows indicate three prominent filamentary ridges, identified by their projected locations as the southeastern (SE), northeastern (NE), and western (W) filaments; the SE filament is the structure analyzed in detail in this work.
    }
    \label{fig:fov}
\end{figure*}

The observations were obtained with the IRAM NOrthern Extended Millimeter Array (NOEMA) on 2024 February 28 and March 3--4 in the A configuration (Project WBP23002; PI Z. Li). They targeted the \(^{12}\mathrm{CO}(1-0)\) line at a rest frequency of 115.2712 GHz with a channel width of 0.65 \kms; the adopted systemic velocity of M33 is \(-179\) \kms \citep{1997ApJ...479..244C}. The 19-pointing mosaic covers the central region to a radius of approximately \(1'\) (Figure~\ref{fig:fov}), or 240 pc at the adopted distance. The NOEMA primary beam at this frequency is 43.7\arcsec, and the total integration time was 8 hr.

The data were calibrated with GILDAS/CLIC\footnote{\url{https://www.iram.fr/IRAMFR/GILDAS}} using the standard NOEMA reduction pipeline. Imaging, deconvolution, and mosaicking were performed with the MAPPING task. We adopted natural weighting and allowed up to 20,000 cleaning iterations until the residual noise was nearly uniform. The final synthesized beam is \(0.91\arcsec \times 0.37\arcsec\) at a position angle of $-2.3^\circ$, corresponding to \(3.7\times1.5\) pc. The pixel size is 0.065\arcsec. The projected baselines span 38.7--1655.9 m, corresponding to a nominal maximum recoverable scale of approximately \(8.3\arcsec\) (\(\sim34\) pc) at 115.27 GHz. Emission on larger scales is therefore expected to be progressively spatially filtered.
The median rms noise over the field (excluding the noisy edge) is $\sim$8.9 mJy beam$^{-1}$ per 0.65 \kms\ channel.

For subsequent analysis, we regridded the cube to 0.15\arcsec\ pixels and 1 \kms\ channels. This sampling retains more than two pixels across the synthesized-beam minor axis and resolves lines with widths of several \kms. We also corrected the radio velocity axis to the laboratory CO(1--0) rest frequency of 115.2712018 GHz. The native cube header used 115.270000 GHz, corresponding to a \(+3.13\) \kms\ shift; all velocities reported below use the corrected LSRK axis.

\subsection{Moment maps}\label{subsec:mom}

We first used the Source Finding Application (SoFiA) software\footnote{\url{https://gitlab.com/SoFiA-Admin/SoFiA-2}} \citep{2015MNRAS.448.1922S,2021MNRAS.506.3962W} to identify significant CO(1--0) emission in the regridded data cube. Although SoFiA was originally developed for H{\sc i} spectral-line source finding, it operates generally on 3D spectral-line data cubes with two spatial axes and one spectral axis and can therefore be used as a blind search for coherent CO emission. A smooth-and-clip finder was applied to the regridded NOEMA cube, with spatial smoothing kernels of 0, 3, 5, and 7 pixels, spectral smoothing kernels of 0, 3, 5, 7, and 15 channels, and a threshold of \(3.5\sigma\). This kernel set is designed to recover unresolved to moderately resolved structures and to cover both narrow and broad CO line components. The detected pixels were grouped into 3D structures with a friends-of-friends linking radius of two pixels in both the spatial and spectral dimensions. We required minimum extents of five spatial pixels and three channels. SoFiA's reliability filter was applied with a reliability threshold of 0.8, followed by a minimum integrated S/N of 5. The resulting SoFiA 3D mask was used for subsequent clump decompositions. 
We label individual SoFiA detections L1, L2, and so forth, using the source identifiers assigned by SoFiA. These labels refer only to structures in the SoFiA segmentation mask and are distinct from the {\sc pycprops} compact clumps used in the statistical analysis below.

\begin{figure*}[htpb!]
    \centering
    \includegraphics[width=\linewidth]{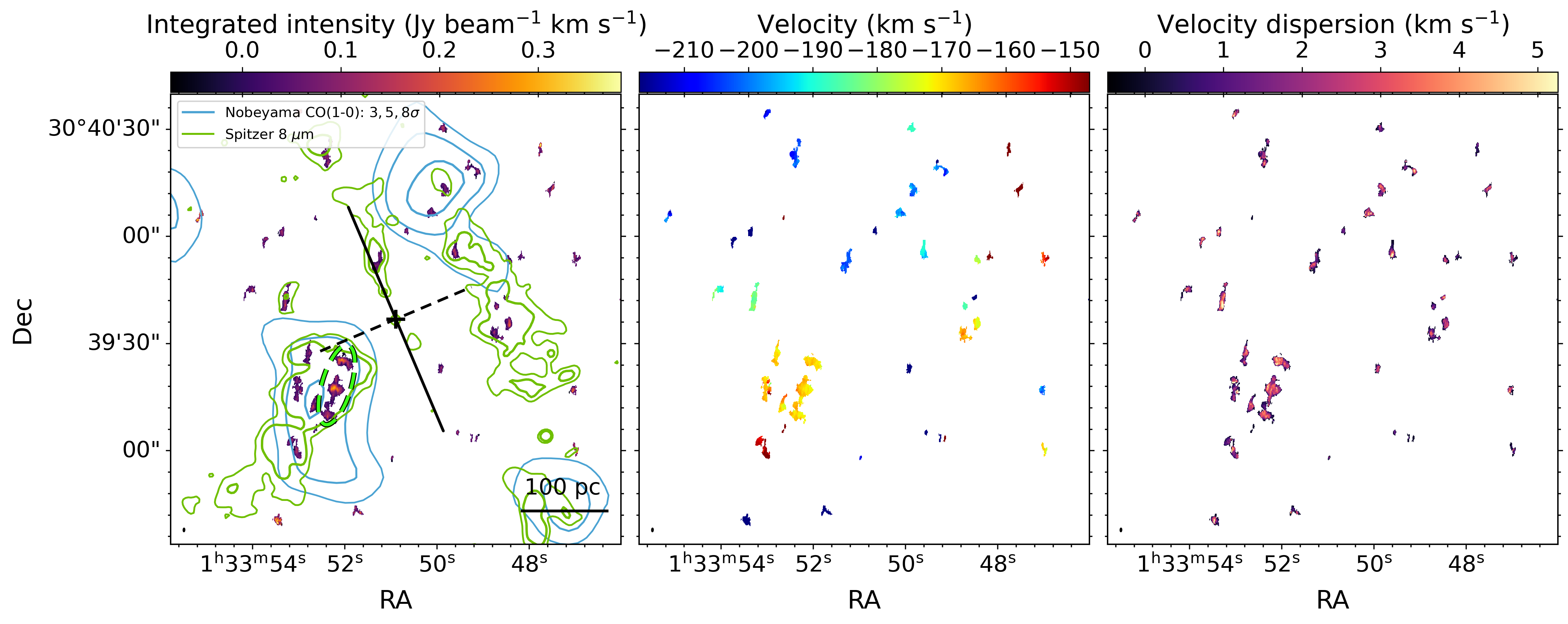}
    \caption{NOEMA CO(1--0) moment maps derived within the SoFiA mask. From left to right, the panels show integrated intensity (m0), intensity-weighted velocity (m1), and velocity dispersion (m2). In the left panel, the black cross marks the adopted M33 center; the solid and dashed black lines show the projected disk major axis (PA \(=23^\circ\) east of north) and minor axis, respectively. The schematic axis of the proposed central stellar bar (PA $\sim 114^\circ$; \citealp{2023ApJ...957....3S}) is approximately aligned in projection with M33's disk minor axis (PA $\sim 113^\circ$).
    Blue contours in the left panel show the Nobeyama CO(1--0) integrated intensity at \(3\sigma\), \(5\sigma\), and \(8\sigma\), where \(\sigma=0.74~{\rm K\,km\,s^{-1}}\); the Nobeyama resolution is 19.3\arcsec. The Spitzer 8~$\mu$m contours are the same as those in Figure~\ref{fig:fov}. The black ellipse at the lower left shows the NOEMA synthesized beam. The dashed green ellipse outlines the SE filament analyzed in Sections~\ref{subsec:offset} and~\ref{subsec:filkin}.} \label{fig:moments}
\end{figure*}

The moment maps were generated within the SoFiA mask. Figure~\ref{fig:moments} shows the integrated intensity in Jy beam$^{-1}$ \kms, intensity-weighted velocity, and velocity dispersion. The total flux within the mask is approximately 40 Jy \kms. A comparison with moment maps generated using a local-rms threshold indicates that SoFiA recovers most of the emission detected above \(3\sigma\) (Appendix~\ref{append:A}). We adopt the SoFiA mask because a local-rms threshold can exclude lower-S/N emission that is coherent in position and velocity, whereas integration over a fixed velocity interval adds noise from line-free channels. The multi-scale three-dimensional SoFiA mask retains statistically reliable coherent emission while limiting this additional noise.

\subsection{Ancillary \texorpdfstring{H$\alpha$}{H-alpha}, CO, and stellar data}

We use H$\alpha$ imaging to compare the distribution of the compact molecular gas with recent massive star formation. The Local Group Galaxy Survey H$\alpha$ map provides galaxy-scale context \citep{2006AJ....131.2478M,2007AJ....134.2474M}, while a continuum-subtracted HST ACS/WFC image (Zhang et al. in prep) constructed from the F658N and F625W bands resolves the ionized gas within the NOEMA field. 
Gaia-aligned HST products typically achieve absolute positional accuracies better than 0.02\arcsec\ \citep{2025DrizzlePac}. For interferometric measurements, the statistical position error is of order the synthesized beam divided by the detection S/N. Empirically, this is $\sim$ 1/10 of the synthesized beam under suitable conditions \citep{1997PASP..109..166C,2002IRAMSchool}. We therefore adopt 0.10\arcsec\ as a conservative empirical NOEMA astrometric uncertainty. 
Adding the HST and NOEMA terms together gives a relative astrometric uncertainty of approximately 0.10\arcsec, corresponding to 0.4 pc at 840 kpc. We also use the PHATTER stellar catalog \citep{10.17909/t9-ksyp-na40,2021ApJS..253...53W} to locate the recent stellar population relative to the CO and H$\alpha$ emission. Following the photometric selection of \citet{2021ApJS..253...53W}, we identify OB-star candidates with reliable UV photometry, $F336W-F475W<0$, and $M_{F475W}<0$.
We also use the Spitzer/IRAC 8~$\mu$m image from the Local Volume Legacy survey \citep{2009ApJ...703..517D,2020LVLdata}.
Lower-resolution Nobeyama CO(1--0) and ACA+IRAM CO(2--1) maps are available for M33 \citep{2011PASJ...63.1171T,2014A&A...567A.118D,2023ApJ...953..164M}, but their angular resolutions and spatial responses differ substantially from those of the NOEMA data ($\sim30$--80 pc compared with a few parsecs). Thus, the Nobeyama and NOEMA data cannot be combined to produce a high-resolution joint image comparable to that of CO(2--1). Consequently, these data are not used for measurements of cloud properties or line ratios. 
We use the Nobeyama CO(1--0) map only to estimate the fraction of large-scale flux recovered after smoothing the NOEMA map to the single-dish resolution.

To quantify the separation between the CO and H$\alpha$ ridges, both images are smoothed to a common 1.5\arcsec\ resolution. We determine the principal axis of the SoFiA-defined filament and construct transverse profiles in independent 8 pc bins along it. Ridge positions are measured from the intensity-weighted centroids; segments whose CO peak is below 10\% of the maximum segment peak are excluded. We estimate the statistical uncertainty on the median separation with 20,000 bootstrap resamplings; the relative astrometric uncertainty is reported separately. 
As a complementary two-dimensional estimate, we apply a normalized cross-correlation method commonly used for astronomical image comparison \citep{2008MNRAS.386..619C}. We shift the H$\alpha$ map over a grid of spatial offsets and calculate its pixel-by-pixel Pearson correlation with the fixed CO map at each position. The shift that maximizes the correlation is adopted as the characteristic displacement between the two emission distributions.

For the spatial comparison, we further restrict this photometrically selected sample to $F475W<22$ ($M_{F475W}\lesssim-2.7$ at the adopted distance). This additional cut emphasizes the most luminous, short-lived candidates and reduces contamination from fainter sources. We measure the shortest projected distance from each source to the significant CO and H$\alpha$ emission. A candidate is considered associated with a tracer when this distance is smaller than 6 pc, approximately the common \(1.5\arcsec\) resolution. 

\section{Results}\label{sec:result}

\subsection{Whole-field molecular-gas distribution}

Figure~\ref{fig:moments} shows compact CO emission throughout the central field, with an integrated flux of approximately 40 Jy \kms\ within the SoFiA mask. The velocity field changes from predominantly blueshifted values in the northwest to redshifted values in the southeast, broadly following the projected disk rotation. Most detected structures have moment-based velocity dispersions of a few \kms, although locally broader profiles occur where spatial or line-of-sight components overlap.

The compact CO emission broadly follows the spiral-like ridges traced by HST H$\alpha$ and Spitzer/IRAC 8~$\mu$m emission \citep{2009ApJ...703..517D} in the central region, but NOEMA resolves these ridges into discrete clumps (Figure~\ref{fig:fov}). Additional isolated clumps lie outside the prominent multi-tracer ridges.

The brightest coherent feature is the southeastern structure outlined in Figure~\ref{fig:moments}, hereafter the SE filament. The Nobeyama CO(1--0) contours from the all-disk survey \citep{2011PASJ...63.1171T} trace broader molecular complexes that are fragmented in the NOEMA map. 

This difference is particularly evident in the SE: Nobeyama detects a diffuse envelope around the molecular complex, whereas NOEMA recovers primarily compact clumps within it (Figure~\ref{fig:moments}). After smoothing the NOEMA map to the 19.3\arcsec\ Nobeyama resolution and applying a common aperture, the smoothed NOEMA map recovers approximately 20--30\% of the CO(1--0) flux over the full field and $\sim 1/3$ in the SE complex. Because convolution cannot restore spatial frequencies missing from the interferometric data, this mismatch is consistent with spatial filtering of extended emission. The reported errors in the Nobeyama intensity-scaling factors are smaller than approximately 20\%, so these recovery fractions should be regarded as approximate \citep{2011PASJ...63.1171T}. Even allowing for this uncertainty, the large flux deficit indicates that the NOEMA map primarily traces the compact, high-surface-brightness component rather than the complete molecular reservoir. This distinction is retained below when masses and virial parameters are interpreted.

\subsection{Compact-clump identification and property estimation}\label{sec:identify}

We characterize compact structures with {\sc pycprops} \citep{2021MNRAS.502.1218R}, using the SoFiA 3D source mask to restrict the decomposition to significant CO emission. {\sc pycprops} identifies the hierarchical emission structure with {\sc astrodendro} and separates local maxima with a watershed algorithm. We require at least 30 voxels per dendrogram structure and a contrast of \(2\sigma\) above the merge level. Local maxima are defined within a neighborhood of two spatial pixels and one spectral channel, and the watershed compactness parameter is 0.7. These choices conservatively separate compact peaks within connected emission.

Following the CPROPS methodology \citep{2006PASP..118..590R}, sizes and linewidths are extrapolated to zero intensity and corrected for the beam and channel response. The effective radius is \(R=1.91(\sigma_{\rm maj}\sigma_{\rm min})^{1/2}\). We adopt a distance of 840 kpc \citep{2001ApJ...553...47F} and a CO-to-H$_2$ conversion factor \(\alpha_{\rm CO}=4.3~M_\odot~({\rm K~km~s^{-1}~pc^2})^{-1}\), including helium \citep{2013ARA&A..51..207B}, to calculate \(M_{\rm mol}=\alpha_{\rm CO}L_{\rm CO}\). {A dust-based study of M33 found a value about twice as high \citep{2017A&A...600A..27G}; we consider this alternative below. Following \citet{2013ApJ...768L...5L}, we define}
\begin{equation}
    \Sigma_{\rm mol}=\frac{M_{\rm mol}}{\pi R^2}
\end{equation}
and
\begin{equation}
    \alpha_{\rm vir}=\frac{5\sigma_v^2R}{aGM_{\rm mol}},
\end{equation}
where \(a=10/9\) is the geometric correction for a spherical density profile \(\rho\propto r^{-1}\). For this definition, \(\alpha_{\rm vir}=1\) corresponds to virial balance in the idealized absence of magnetic and surface-pressure terms, while \(\alpha_{\rm vir}=2\) is commonly used as the approximate boundary for a marginally bound isolated cloud. {The CO linewidth includes thermal broadening but is dominated by nonthermal motions. Random, approximately isotropic motions may provide pressure support, whereas coherent inflow, expansion, or shear may not \citep{1992ApJ...399..551M}. Thus, \(\alpha_{\rm vir}\), calculated from the full linewidth, does not distinguish between supportive random motions and coherent motions.} Statistical uncertainties for individual clumps are estimated with 1000 bootstrap resamplings. We characterize each clump-property distribution using its median and observed 16th--84th percentile range. These percentiles describe the intrinsic spread of the measured values and are not uncertainties on the median.

Motivated by the classical Milky Way molecular-cloud relation, we fit size--linewidth normalizations \(\sigma_v=A R^{1/2}\) \citep{1987ApJ...319..730S} for the clumps by inverse-variance weighting. A slope of \(1/2\) is also expected for clouds with similar surface densities and virial parameters, although the normalization \(A\) may vary with both quantities \citep{2009ApJ...699.1092H}. Fixing the slope is appropriate for the small subsamples and permits a direct comparison of their size--linewidth normalizations.

Because the catalog depends on the SoFiA mask and decomposition parameters, we treat the identified objects as compact clumps for comparative analysis rather than as a unique segmentation of physical clouds. The statistical sample is further restricted to clumps with finite, positive deconvolved properties and directly measured bootstrap uncertainties.

\begin{figure*}[htpb!]
    \centering
    \includegraphics[width=0.90\linewidth]{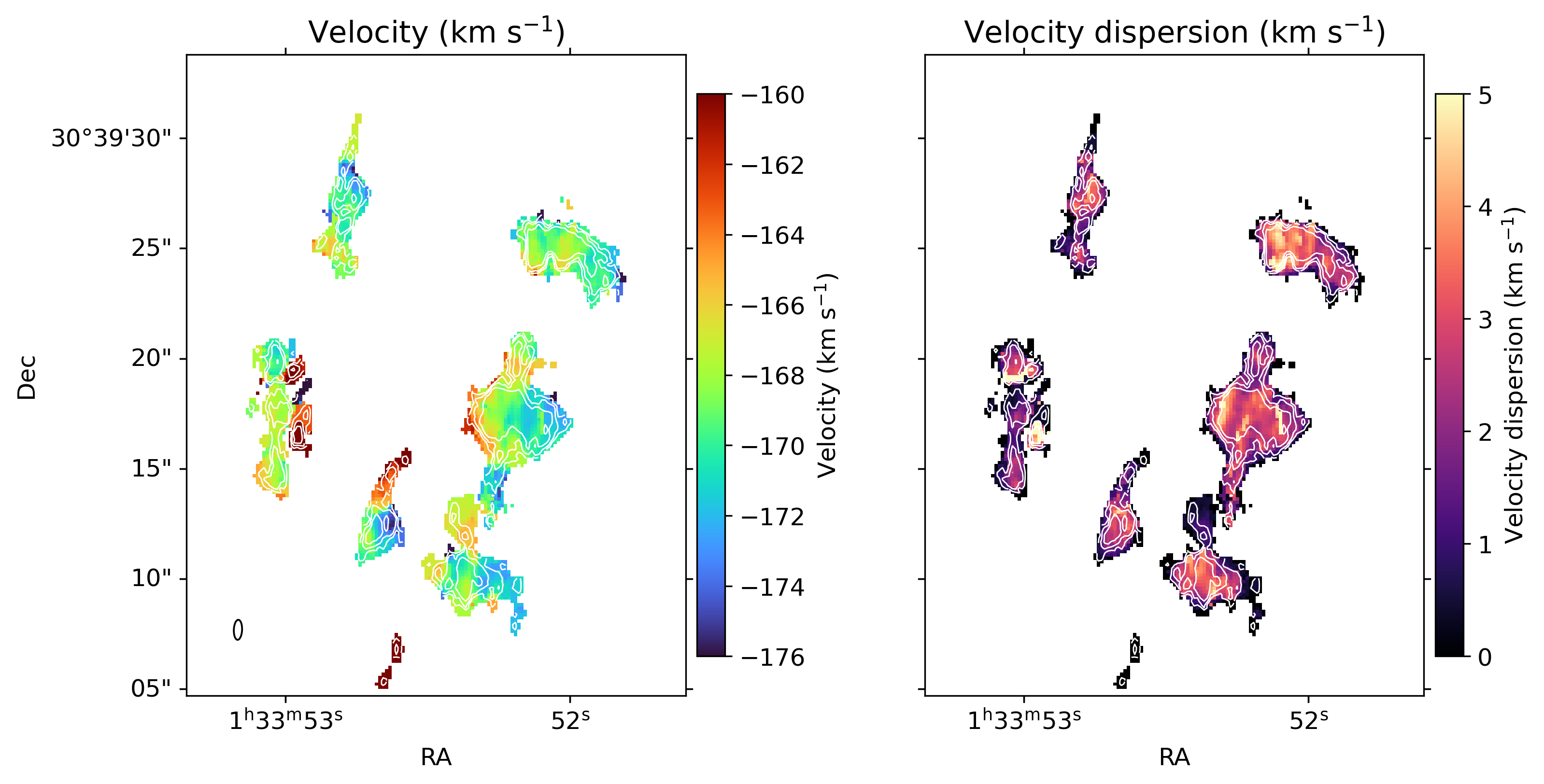}
    \caption{{Zoomed NOEMA CO(1--0) kinematic maps of the SE region. The left panel shows intensity-weighted velocity and the right panel velocity dispersion. White contours mark integrated intensities of 0.033, 0.061, and 0.108 Jy beam$^{-1}$ \kms. The synthesized beam is shown in the lower left of the velocity panel.}}
    \label{fig:se_moments_zoom}
\end{figure*}

\subsection{The SE filament and recent star formation}\label{subsec:offset}

The SE filament extends approximately 76 pc along its principal axis and 23 pc across it in projection, as measured from the full extent of the SoFiA mask.

\begin{figure*}[htpb!]
    \centering
    \includegraphics[width=0.5\linewidth]{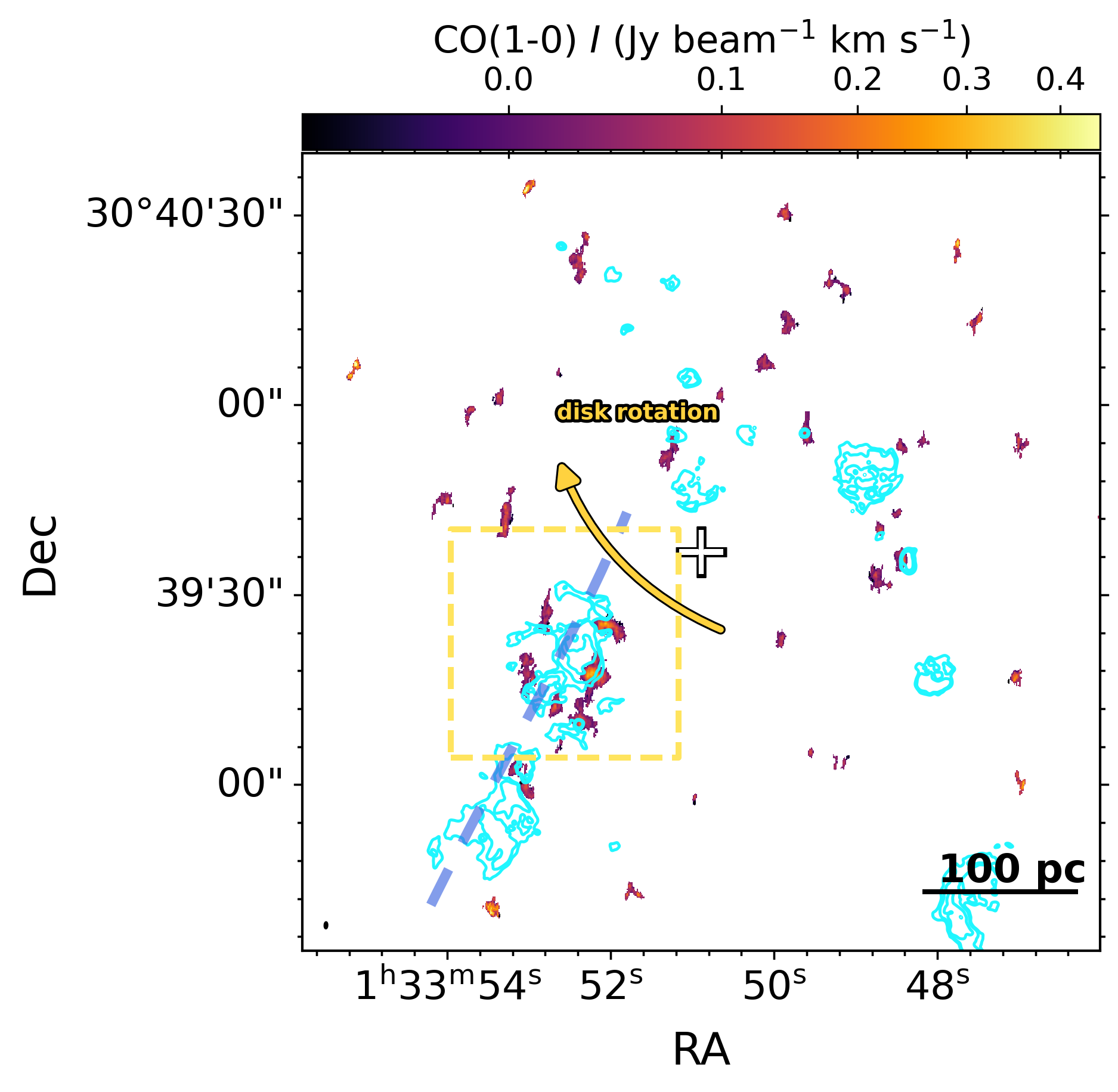}
    \includegraphics[width=0.49\linewidth]{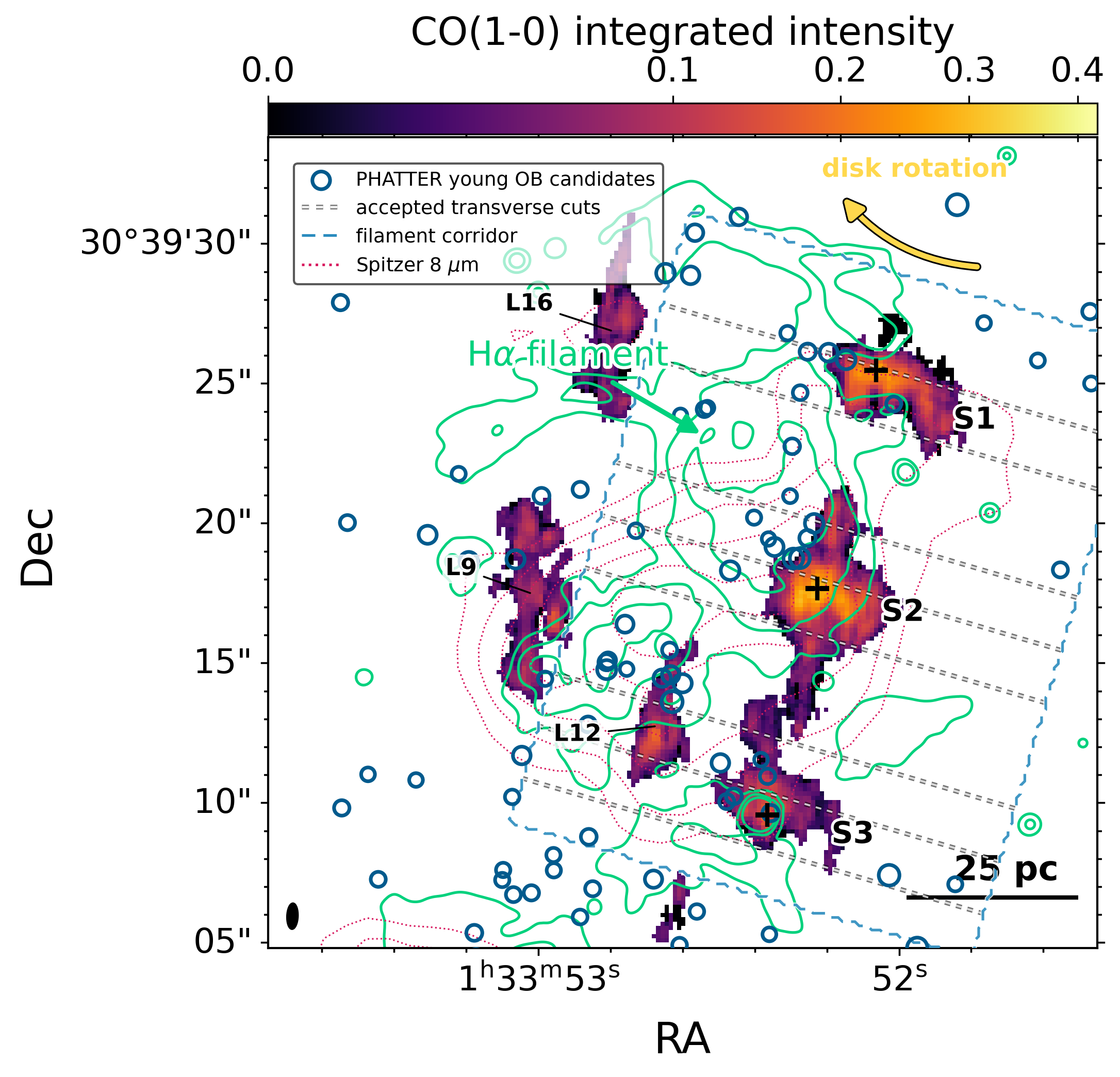}
    \caption{\textit{Left:} NOEMA CO(1--0) integrated intensity with HST H$\alpha$ contours (cyan). The yellow rectangle marks the region enlarged in the right panel. The cross marks the adopted M33 center, and the curved yellow arrow shows the projected clockwise disk rotation \citep{2018MNRAS.479.2505K}. The semi-transparent blue dashed curve is a schematic inner-arm guide through the southeastern multi-tracer ridge, not a fitted or previously cataloged spiral-arm locus. \textit{Right:} Enlarged view of the SE filament. Colors show CO(1--0), green contours show H$\alpha$, and blue circles mark young OB candidates from PHATTER \citep{2021ApJS..253...53W}. Circle size scales with F475W brightness. The curved yellow arrow indicates the projected disk rotation, the blue dashed outline marks the filament corridor used in the analysis, and the short gray dashed lines show the accepted transverse cuts. The three bright knots S1--S3 are marked with crosses. L9, L12, and L16 label neighboring SoFiA structures; L12 and L9 are discussed in Appendix~\ref{append:spectra}, while L16 is excluded based on its distinct morphology.
    }
    \label{fig:offset}
\end{figure*}

NOEMA resolves the ridge into three principal knots, S1--S3 (Figure~\ref{fig:offset}); S2 is the brightest. Integrating the CO(1--0) emission in the moment-zero map over this region gives \(L_{\rm CO}=1.51\times10^4~{\rm K\,km\,s^{-1}\,pc^2}\), corresponding to \(M_{\rm mol}=6.6\times10^4~M_\odot\) for the adopted \(\alpha_{\rm CO}\).

This value represents the compact molecular component and should be considered a lower limit to the total filament mass because the NOEMA data spatially filter extended emission. 

The HST H$\alpha$ image shows an elongated chain of ionized-gas peaks adjacent to the CO ridge (Figure~\ref{fig:offset}). The relative location is ordered over most of the filament: H$\alpha$ lies primarily east and north of the principal CO ridge rather than surrounding each CO peak symmetrically. Six of the eight reliable transverse cuts place H$\alpha$ on the same side of CO. Their median signed H$\alpha$-minus-CO ridge separation is \(10.0^{+0.9}_{-1.4}\) pc, where the interval is the bootstrap 16th--84th percentile uncertainty on the median. The approximate relative-astrometry uncertainty is 0.4 pc, much smaller than the measured separation. The independent two-dimensional cross-correlation rises from 0.21 at zero lag to 0.56 at its maximum, corresponding to a projected displacement of 16.0 pc toward the east and north. The increase after shifting also shows that the ionized-gas ridge approximately follows the elongated shape of the molecular ridge. The two displacement estimates differ because the transverse profiles follow the local filament axis, whereas the cross-correlation weights the full two-dimensional morphology.

There are 46 bright young OB candidates with \(F475W<22\) within the filament corridor (Figure~\ref{fig:offset}). Using a 6 pc matching radius, 39 candidates are associated with high H$\alpha$ emission and 19 with compact CO. All 19 CO-associated candidates are also associated with H$\alpha$; an additional 20 are associated only with H$\alpha$, while 7 are associated with neither tracer. No candidate is associated with CO alone. For comparison, regions within 6 pc of H$\alpha$ and CO peaks cover 66\% and 36\% of the filament corridor, respectively. The stellar catalog therefore indicates a closer association of the young stellar population with the ionized gas than with the compact molecular emission. The OB candidates appear concentrated near the S2 CO--H$\alpha$ interface (Figure~\ref{fig:offset}), although the present counts do not establish a distinct enhancement specifically at the interface.

Two neighboring compact CO structures, L12 and L9, lie near apparently connected Spitzer 8~$\mu$m emission and warrant separate consideration (Figure~\ref{fig:offset}; Appendix~\ref{append:spectra}). L16 is clearly separated from the SE filament in the 8~$\mu$m map and is excluded from the subsequent analysis. L12 lies adjacent to S3 and near the southern H{\sc ii} region. Its fitted velocity centroid, \(-169.3\) \kms, is compatible with that of S3, \(-168.6\) \kms, although its broad and irregular spectrum and the absence of a continuous CO bridge prevent a secure physical association. L12 could trace an extension of the S3 environment, gas compressed or disturbed near the H{\sc ii} region, or an unrelated projected structure. L9 is more distinct from the SE filament: its fitted centroid is \(-167.1\) \kms\ and its linewidth is narrower than those of S1--S3. A dedicated S3--L12--L9 position-velocity diagram (PVD) also contains a 13 pc interval without SoFiA-selected emission between L12 and L9 (Figure~\ref{fig:pvd}). We therefore exclude L9 from the fiducial SE filament. Its proximity may be a chance projection, although a response to the same large-scale compression or to local feedback cannot be excluded.

\subsection{Kinematics of the SE filament}\label{subsec:filkin}

The velocity field (m1) changes little along the full SE filament. {The enlarged moment maps in Figure~\ref{fig:se_moments_zoom} and the native-resolution channel maps in Appendix~\ref{append:channels} show that S1--S3 remains spatially coherent across the main line channels.} Intensity-weighted linear fits to the m1 values give an along-axis gradient of \(0.005\pm0.002~{\rm km~s^{-1}~pc^{-1}}\), corresponding to $<$ 0.4 \kms\ across its 76 pc projected extent. In contrast, the fitted cross-axis gradient is \(-0.259\pm0.009~{\rm km~s^{-1}~pc^{-1}}\), corresponding to several \kms\ across the bright ridge. The formal fit errors describe the scatter of the m1 samples and do not include uncertainties from mask selection or multiple line components.
Using the H$\alpha$ rotation-curve model derived from Fabry--P\'erot interferometric observations \citep{2015MNRAS.449.4048K}, we estimate that projected disk rotation accounts for only \(\sim20\)--30\% of the measured velocity change across the SE filament.

\begin{figure*}[htpb!]
    \centering
    \includegraphics[width=0.92\linewidth]{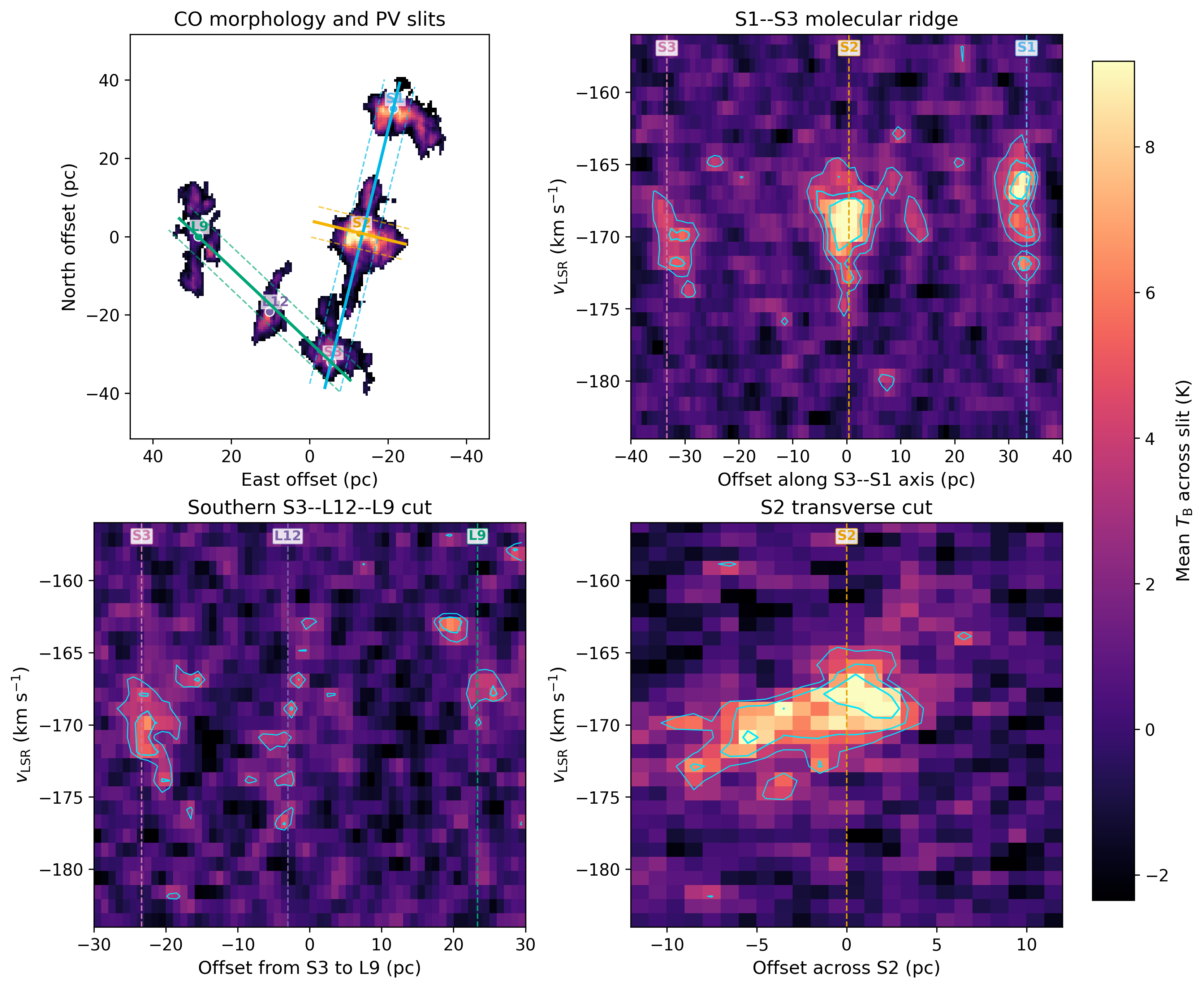}
    \caption{Position--velocity structure of the SE filament and neighboring CO emission. {From upper left to lower right, the panels show} the integrated-intensity map with the three adopted slits, the PVD along the S1--S3 axis, the PVD along the candidate S3--L12--L9 sequence, and the transverse PVD through S2. Dashed lines in the map indicate the slit widths. Colors show the mean CO brightness temperature in the unmasked cube, and cyan contours mark emission at 3, 5, and 8 times the PVD rms. {The displayed velocity range is $-184$ to $-156$ \kms.} The S1--S3 PVD is dominated by the three compact knots, the S3--L12--L9 cut lacks a continuous bridge, and the S2 transverse PVD shows the clearest organized velocity shift.}
    \label{fig:pvd}
\end{figure*}

The PVDs in Figure~\ref{fig:pvd} clarify this kinematic structure. Along the filament, the emission is separated into the S1--S3 concentrations with modest centroid changes between them. The S3--L12--L9 cut instead shows discrete emission concentrations without a continuous position--velocity bridge, supporting our decision not to extend the fiducial filament through L9. Across S2, the emission shifts coherently in velocity, although it does not form a single straight ridge at every position. We therefore interpret the S2 PVD as evidence for resolved transverse kinematic structure, not as a secure detection of two colliding velocity components. Around S2, the brightest compact CO emission lies toward the H$\alpha$-facing edge, where the PVD shows the clearest transverse velocity change.

\subsection{Whole-field compact-clump properties}\label{sec:clumpprop}

The statistical catalog contains 23 compact clumps: 6 associated with the SE filament and 17 elsewhere in the field. {Their positions and measured properties are tabulated in Appendix~\ref{append:catalog}.} Their radii range from 1.4 to 5.4 pc, with a median of 2.67 pc, and their velocity dispersions range from 0.57 to 4.48 \kms, with a median of 2.15 \kms. These dimensions correspond primarily to molecular clumps and small cloud substructures rather than entire GMCs or subparsec dense cores \citep{2000prpl.conf...97W,2007ARA&A..45..339B}.

\begin{figure*}[htpb!]
    \centering
    \includegraphics[width=\linewidth]{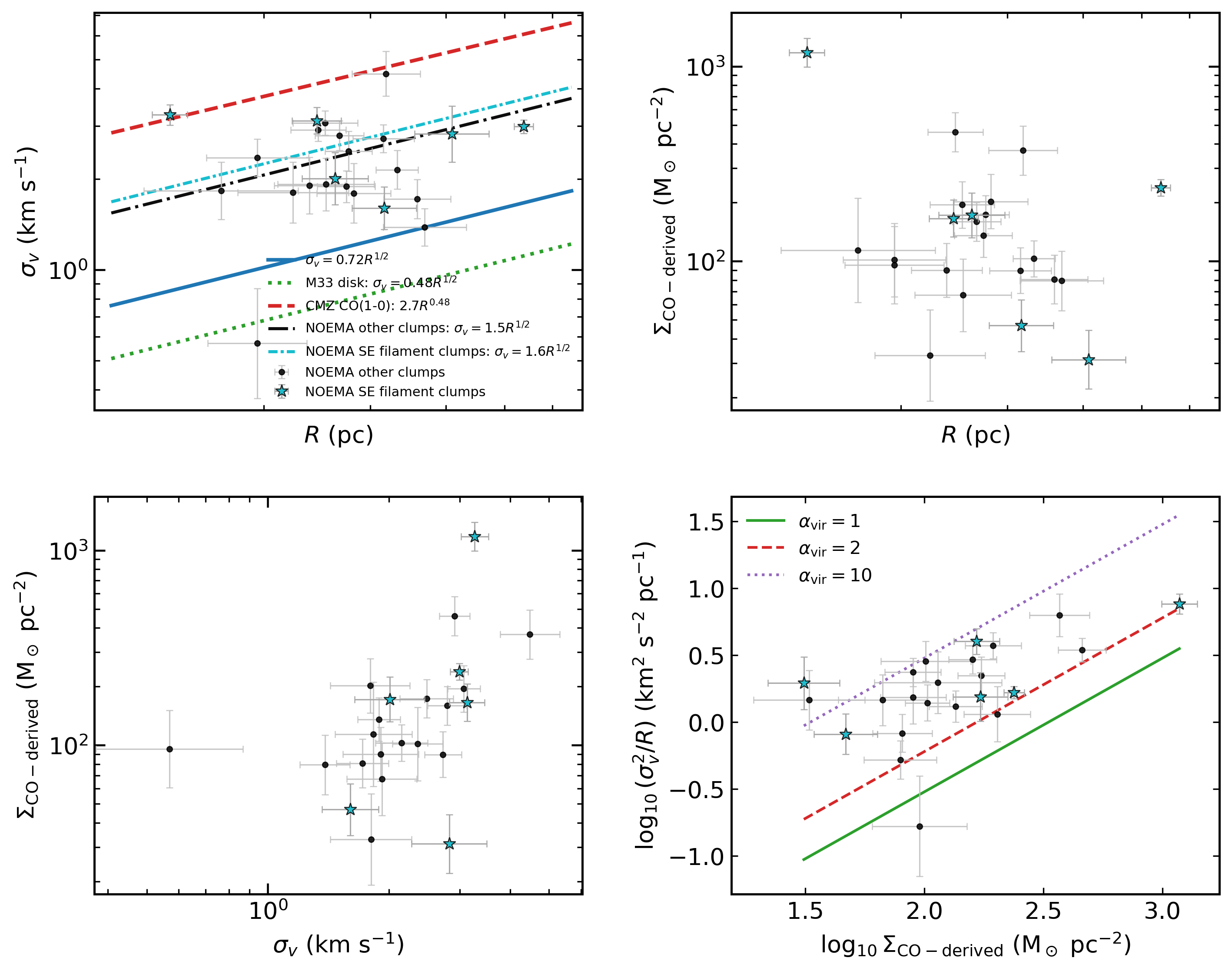}
\caption{Properties of NOEMA CO(1--0) compact clumps identified with {\sc pycprops}. Cyan stars mark clumps associated with the SE filament, and black circles mark the other clumps in the field. \textit{Upper left:} Radius--velocity dispersion relation, compared with the Milky Way disk relation \citep{1987ApJ...319..730S}, {the M33 disk-wide relation \citep{2023ApJ...953..164M},} and the CMZ fit \citep{1998ApJS..118..455O}; dash-dotted lines show fits with a fixed \(R^{1/2}\) slope to the two NOEMA subsamples. \textit{Upper right:} Radius--surface density relation. \textit{Lower left:} Surface density--velocity dispersion relation. \textit{Lower right:} Size--linewidth--surface density relation, with lines corresponding to virial parameters of 1, 2, and 10.}
    \label{fig:correlation}
\end{figure*}

At a given size, most NOEMA clumps lie above the Milky Way disk size--linewidth relation and below the Milky Way CMZ relation \citep[Figure~\ref{fig:correlation};][]{1987ApJ...319..730S,1998ApJS..118..455O}. {For comparison, Figure~\ref{fig:correlation} includes the disk-wide M33 relation, \(\sigma_v=0.48R^{1/2}\), measured at approximately 30 pc resolution by \citet{2023ApJ...953..164M}.} Fixed-slope fits of the form \(\sigma_v = A R^{1/2}\) give \(A = 1.59\pm0.27~{\rm km~s^{-1}~pc^{-1/2}}\) for the SE-filament clumps and \(A=1.46\pm0.12~{\rm km~s^{-1}~pc^{-1/2}}\) for the other clumps. {These normalizations are about three times the disk-wide value, although the physical resolution and decomposition methods differ.} The similar normalizations and linewidth distributions provide no evidence for a clump-scale linewidth enhancement in the SE filament.

The median CO-derived mass and surface density are \(3.27\times10^3~M_\odot\) and \(114~M_\odot~{\rm pc^{-2}}\), respectively. The surface densities span 31--1178 \(M_\odot~{\rm pc^{-2}}\), with the observed 16th--84th percentile range extending from 73 to 220 \(M_\odot~{\rm pc^{-2}}\). The median apparent virial parameter is 5.67, with an observed 16th--84th percentile range of 2.26--8.38. The SE and non-SE medians are 4.39 and 5.67, respectively, and their distributions show no clear separation.
{Adopting the roughly twice-Galactic \(\alpha_{\rm CO}\) reported by \citet{2017A&A...600A..27G} would double the masses and surface densities and halve the apparent virial parameters. The full-sample medians would then be \(6.54\times10^3~M_\odot\), \(228~M_\odot~{\rm pc^{-2}}\), and \(\alpha_{\rm vir}=2.84\), respectively. Radii, linewidths, the CO--H$\alpha$ geometry, and the SE--field comparison would be unchanged.}
These values apply only to the compact flux recovered by NOEMA. Spatial filtering can alter the measured mass, radius, and velocity dispersion; because \(\alpha_{\rm vir}\propto\sigma_v^2R/M\), the direction and magnitude of the resulting bias cannot be determined a priori. The measured values should therefore be regarded as apparent virial parameters of the NOEMA-selected compact structures.

\section{Discussion}\label{sec:discuss}

\subsection{An ordered gas--star-formation sequence}

Three observational findings motivate our interpretation. First, the principal compact CO ridge forms a 76 pc elongated structure rather than a shell centered on an H{\sc ii} region. Second, the adjacent H$\alpha$ ridge lies preferentially on one side of this ridge, with a median projected separation of 10 pc, and the PHATTER candidates follow H$\alpha$ more closely than compact CO. Third, the resolved velocity change is predominantly across the molecular ridge. L12 may be related to the southern S3 environment, but the current data do not establish a continuous connection, while L9 is spectrally distinct and lacks a position--velocity bridge to the filament. The central observational result is therefore the asymmetric concentration of the S1--S3 molecular ridge relative to H$\alpha$, together with its transverse kinematics. This combination is consistent with compression of molecular gas followed by massive star formation and exposure of the ionized regions.

The SE filament belongs to a broader class of spatially ordered gas--star-formation structures. CO--H$\alpha$ offsets have been measured across the spiral arms of several nearby galaxies, although some systems show ambiguous or negligible offsets \citep{2009ApJ...697.1870E}. In M51, the predominantly ordered CO--H$\alpha$ offsets were interpreted as evidence that gas flows through the arms, while their substantial scatter and sensitivity to the adopted gas tracer argued against treating any single offset as proof of a stationary spiral pattern \citep{2013ApJ...763...94L}. At $\sim$100--150 pc resolution, the Physics at High Angular resolution in Nearby GalaxieS (PHANGS) sample of nearby star-forming galaxies likewise shows distinct CO-bright, overlapping CO--H$\alpha$, and H$\alpha$-bright phases whose relative distributions depend on galaxy properties and local galactic structure \citep{2022ApJ...927....9P}. Among 24 PHANGS galaxies with well-defined spiral structure, \citet{2025A&A...701A.183Q} found positive mean CO--H$\alpha$ offsets in 58\% of the sample, but only 17\% showed the radial trend expected for a single quasi-stationary density wave. The substantial scatter indicates that such offsets do not uniquely identify a coherent spiral shock.
The few-parsec-resolution NOEMA data resolve such spatial ordering across an individual molecular ridge, substantially below the physical scale of these galaxy-wide comparisons.

The closest resolved comparison is GMC-16 in the disk of M33. At 1--2 pc resolution, \citet{2020ApJ...896...36T} identified molecular filaments 50--70 pc long and a 10--20 pc projected offset between the CO filaments and neighboring H{\sc ii} regions. The CO data also revealed velocity differences of several \kms among molecular components. The SE filament has a comparable length, CO--H$\alpha$ separation, and transverse velocity range, placing it within the broader population of large molecular structures shaped by galactic-scale gas flows. Unlike the observations of GMC-16, however, the NOEMA data recover only the compact CO(1--0) component and show no clearly separated velocity components. Moreover, the local pattern speed and gravitational forcing are not constrained. Taken together, the morphology, transverse kinematics, and preferential placement of H$\alpha$ relative to the principal CO ridge make large-scale compression a plausible common interpretation. The present data do not determine whether that compression is produced by a classical spiral shock, bar-driven flow, or a transient spiral feature.

\subsection{Spiral compression and alternative interpretations}

Published classifications of M33's spiral structure primarily trace optical and stellar arm segments across the larger disk \citep{1980ApJS...44..319H,2017MNRAS.464.2103J}. Although more recent work suggests underlying old-stellar arms and a possible central bar \citep{2023ApJ...957....3S}, these studies do not define an unambiguous arm locus or dynamically constrained spiral pattern through the inner \(\sim500\) pc sampled by NOEMA. We therefore describe the central multi-tracer features as spiral-like ridges and do not assume that the SE filament lies within an established spiral arm.

Resolved stellar populations have been interpreted as revealing a central bar extending over approximately 1 kpc and two underlying old-stellar arms in M33 \citep{2023ApJ...957....3S}. The NOEMA field lies within this proposed central non-axisymmetric structure, so bar-driven or spiral-driven compression may both contribute to the observed gas morphology. The proposed stellar bar is oriented approximately along the minor axis of M33, but neither its local forcing nor its pattern speed is measured at the SE filament; a bar contribution therefore cannot be quantified.
Relative to the projected clockwise disk rotation in Figure~\ref{fig:offset}, H$\alpha$ lies mainly beyond the principal CO ridge. However, identifying this side as dynamically downstream requires knowing the spiral pattern speed, which has not been measured here. We therefore cannot interpret the CO--H$\alpha$ separation as a unique time delay or confirm that the schematic curve traces a spiral arm.

In a quasi-stationary spiral-shock scenario, gas entering the compression zone may be assembled into denser molecular structures; if collapse follows, massive star formation is expected downstream after a finite delay. 
A transient, approximately co-rotating arm can also assemble molecular gas through converging flows, but this picture does not generally predict that the younger-star tracer will remain on the same side of the molecular ridge \citep{2016MNRAS.460.2472B}. In the SE filament, the elongated S1--S3 ridge, common-sided CO--H$\alpha$ offset in six of eight independent cuts, and predominantly transverse velocity change are consistent with a locally asymmetric compression geometry. However, this ordering could also reflect molecular gas reshaped by feedback from a pre-existing massive-star-forming region.
The ambiguous association of L12 and the separate L9 feature show that the surrounding molecular distribution is more complex than a single monotonic sequence, but neither feature defines a coherent second ridge in the current data. The two cuts without the common-sided offset lie near S3, where H$\alpha$ is associated with a separate southern H{\sc ii} region rather than the elongated H$\alpha$ ridge adjacent to S1--S2. This local morphology may reflect projection or an additional star-forming region and is not used to define the principal gas--star-formation sequence.
The 76 pc S1--S3 structure therefore exhibits an ordered arrangement consistent with compression preceding star formation, but it does not establish that chronology or uniquely identify the driving mechanism.

The observed transverse kinematics provide an order-of-magnitude characteristic timescale. The inverse of the fitted cross-filament velocity gradient (\(\sim -0.26~{\rm km~s^{-1}~pc^{-1}}\)) gives
\[
t_{\rm kin}\equiv\left|\nabla_\perp v_{\rm los}\right|^{-1}
\approx 3.8~{\rm Myr}.
\]
Assuming that both the filament and the transverse motion lie in the disk of M33 gives a similar deprojected characteristic timescale of $\sim 4$ Myr. 
This is comparable to the \(4.3^{+0.1}_{-0.3}\) Myr timescale of continuum-subtracted H$\alpha$ emission \citep{2020MNRAS.498..235H} and to the 1--5 Myr interval over which exposed massive stars coexist with, and rapidly disperse, their parent molecular gas in nearby galaxies \citep{2020SSRv..216...50C}. 
It is also similar to the S1--S3 aperture-scale crossing times of 4--5 Myr and of the same order as their approximate free-fall times of 4--8 Myr, suggesting that the transverse structure can evolve within roughly one knot-scale dynamical time. 

More broadly, CO--H$\alpha$ offset measurements in nearby spiral galaxies infer star-formation timescales comparable to molecular-cloud free-fall times, although clear offsets are recovered in only a subset of systems \citep{2009ApJ...697.1870E}. The numerical agreement of these few-Myr scales supports the plausibility of a rapid molecular-to-ionized sequence in the SE filament, but does not by itself establish a causal delay.

Around S2, the brightest compact CO emission lies toward the H$\alpha$-facing edge, where the transverse velocity change is clearest. This spatial coincidence is suggestive of localized compression at the molecular--ionized-gas interface. However, because CO(1--0) brightness depends on excitation, linewidth, and beam filling as well as column density, it cannot be interpreted uniquely as an increase in molecular-gas surface density.

This timescale comparison is suggestive rather than a direct measurement of the delay after spiral compression. The transverse gradient is measured only along the line of sight and may include internal motions, shear, or feedback, while both the true separation and ridge width are affected by projection. In particular, \(t_{\rm kin}\) is not the spiral-arm crossing time \(W_{\rm arm}/v_{\rm rel}\), which requires the gas velocity relative to the local spiral pattern. Measuring that quantity would require a constrained pattern speed and a model of the three-dimensional non-circular flow, which is beyond the scope of the current paper.

Feedback provides another viable contribution: ionization and winds can remove CO near young stars and leave molecular gas on the edge of an H{\sc ii} region. {Molecular hydrogen can also survive where CO is weak or dissociated \citep{2026SCPMA..6969513X}.} A simple expanding-shell picture is disfavored by the elongated, one-sided morphology and by the absence of a clear split PVD, but an incomplete or density-bounded shell cannot be excluded. Shear or superposed clouds could likewise generate the transverse velocity gradient without triggering the observed star formation.

Small-scale gas--star offsets also depend on tracer lifetimes, stellar drift, extinction, and stochastic sampling \citep{2014MNRAS.439.3239K,2020MNRAS.498..235H}. The source counts here establish that the selected young stars are more closely associated with H$\alpha$ than with compact CO. Several luminous OB-star candidates east of S2 lie within the adjacent H$\alpha$ emission and could power an H{\sc ii} region bordering the molecular ridge. Their distribution does not demonstrate that they formed at the present boundary or that compression produced a stellar excess there. The principal SE ridge therefore exhibits an ordered projected arrangement of molecular gas, ionized gas, and young stars that is compatible with earlier gas compression, although the present CO--H$\alpha$ boundary may also have been reshaped by feedback.

Determining the driving mechanism is beyond the scope of the present data. A combined compact-configuration NOEMA and IRAM 30 m CO(1--0) cube could recover the filtered emission and test whether S1--S3 and L12 are connected by a continuous molecular component. 
Multi-transition CO observations at comparable resolution could test whether the bright interface reflects enhanced excitation as well as gas concentration, while dense-gas tracers could determine whether the compression extends to denser material.
Finally, resolved ionized-gas kinematics together with a model of the central stellar potential would help separate compression, shear, and feedback and constrain the gas motion in the local pattern frame.

\subsection{Compact clumps as environmental context}

The whole-field catalog indicates that elevated clump-scale linewidths are not unique to the SE filament: the two subsamples have similar \(R^{1/2}\) normalizations, and their linewidth and virial-parameter distributions show no clear separation. Because the NOEMA catalog preferentially selects compact, high-surface-brightness emission, this comparison applies only to the compact molecular component and may not capture environmental differences in the more extended gas. The SE filament's cross-filament velocity gradient instead represents an organized kinematic feature on tens-of-parsec scales.

The normalization \(A\approx1.5\)--1.6 \(\mathrm{km~s^{-1}~pc^{-1/2}}\) is about twice the classical Milky Way disk value. In the LMC, \citet{2019ApJ...885...50W} found that velocity dispersion at fixed size increases with both molecular-gas surface density and star-formation activity. The central M33 clumps may likewise reflect a combination of cloud-scale gravity, converging motions, and feedback. Most remain below the much larger linewidths characteristic of Milky Way CMZ clouds, consistent with the shallower central potential of M33. A qualitative comparison within M33 points in the same direction. At GMC scales, \citet{2018A&A...612A..51B} found that both linewidth and size-normalized linewidth decrease with galactocentric radius. The elevated normalization of our central clumps is therefore consistent with stronger internal motions toward the center of M33. However, a quantitative comparison is precluded by the different spatial resolutions, CO transitions, and structure-identification methods.

For qualitative context, the apparent median \(\alpha_{\rm vir}=5.67\) is higher than the typical values of \(\sim1.5\)--3 measured for GMCs in the M33 disk, the Milky Way disk, and most nearby-galaxy disks \citep{2009ApJ...699.1092H,2018ApJ...860..172S,2021MNRAS.502.1218R,2024PASJ...76.1098K}. It is numerically closer to the elevated values found for compact clumps in the Milky Way CMZ \citep{2022ApJ...929...34M}. More generally, cloud-scale surveys find enhanced linewidths and, in some cases, higher apparent virial parameters toward galaxy centers, particularly in barred systems \citep{2018ApJ...860..172S,2020ApJ...901L...8S}. These comparisons are qualitative because the measurements differ in physical scale, decomposition method, and sensitivity to extended emission.

More importantly, NOEMA recovers only 20--30\% of the full-field single-dish flux, whereas the {\sc pycprops} extrapolation raises individual clump luminosities by a median of only 10\%. The dominant missing component is therefore spatially extended emission, not a small correction immediately below the clump boundary. If part of that filtered mass belongs to the same physical structures, their true surface densities are higher and virial parameters lower. Uncertainty in \(\alpha_{\rm CO}\), cloud geometry, magnetic fields, external pressure, and unresolved velocity gradients adds further systematic uncertainty \citep{2013ARA&A..51..207B}. The apparent virial parameters should therefore not be interpreted as evidence that the parent molecular structures are globally unbound. The clump analysis is useful as environmental context and for the relative SE-versus-field comparison, but not as the central dynamical argument of this work.

\section{Summary}\label{sec:sum}

We use NOEMA CO(1--0) observations of the inner $\sim$500 pc of M33 to resolve its compact molecular component at a spatial resolution of a few parsecs. Our main findings are:
\begin{itemize}
    \item The principal coherent feature is a 76 pc SE molecular filament containing the three bright CO knots S1--S3. Across eight reliable transverse cuts, the H$\alpha$ ridge is displaced preferentially to one side of the principal CO ridge by a median \(10.0^{+0.9}_{-1.4}\) pc. The gas shows a resolved cross-filament velocity gradient but little velocity change along the filament. The combined morphology and kinematics are consistent with locally asymmetric compression followed by massive star formation, but do not uniquely identify a spiral shock.

    \item The statistical catalog contains 23 compact clumps with directly measured property uncertainties, including 6 on the SE filament. Their median radius, velocity dispersion, surface density, and apparent virial parameter are 2.67 pc, 2.15 \kms, \(114~M_\odot~{\rm pc^{-2}}\), and 5.67, respectively. The linewidth and virial-parameter distributions of the SE and non-SE clumps show no clear separation. Because NOEMA recovers only a small fraction of the extended CO flux, these measurements characterize the compact molecular component and do not imply that the parent complexes are globally unbound.
\end{itemize}

{Future compact-configuration NOEMA observations combined with IRAM 30 m data could recover the extended CO reservoir. Targeted ALMA observations of higher-\(J\) CO and dense-gas tracers at comparable resolution could test whether the bright interfaces reflect enhanced excitation or a genuine increase in the dense-gas fraction. Together with matched ionized-gas kinematics and a model of the central stellar potential, these data would help distinguish spiral compression, transient converging flows, and feedback.}

\begin{acknowledgments}

This work was supported by the National Key Research and Development Program of China (No. 2022YFA1605000) and the National Natural Science Foundation of China (grant 12225302). Z.N.L. acknowledges support from the East Asian Core Observatories Association Fellowship.

This work is based on observations carried out under project number WBP23002 with the IRAM NOEMA Interferometer. IRAM is supported by INSU/CNRS (France), MPG (Germany), and IGN (Spain). We thank the IRAM staff for their support during the observations and data reduction.

\end{acknowledgments}

\facilities{NOEMA}

\software{Astropy \citep{2013A&A...558A..33A,2018AJ....156..123A,2022ApJ...935..167A},
GILDAS \citep{2005sf2a.conf..721P},
SoFiA \citep{2015MNRAS.448.1922S,2021MNRAS.506.3962W},
{\sc pycprops} \citep{2021MNRAS.502.1218R},
{\sc astrodendro} (\url{https://dendrograms.readthedocs.io})}

\appendix

\section{Comparison with local-RMS moment maps}\label{append:A}

To assess the robustness of the SoFiA mask, we independently construct integrated-intensity maps using pixel-dependent rms estimates. Before integration, we retain emission above either \(3\sigma\) or \(2\sigma\) in at least two consecutive native channels, thereby rejecting isolated spectral peaks. Figure~\ref{fig:compare} compares these maps with the SoFiA result. The full-field positive flux of $\sim 40$ Jy \kms in the \(3\sigma\) map is numerically comparable to that obtained with the SoFiA mask , but the spatial distributions are not expected to match pixel by pixel. The direct clip requires a native-resolution spatial pixel to exceed the threshold in consecutive channels, whereas SoFiA first searches spatially and spectrally smoothed cubes and then applies the accepted 3D mask to the unsmoothed data. It can therefore recover a resolved, coherent line whose individual native voxels are mostly below \(3\sigma\). Conversely, positive clipping over a large cube accumulates many sparse threshold crossings outside coherent source footprints. The \(2\sigma\) map  makes this latter effect particularly clear, containing widespread low-level structure  dominated by noise.

\renewcommand{\thefigure}{A\arabic{figure}}
\renewcommand{\theHfigure}{A\arabic{figure}}
\setcounter{figure}{0}

\begin{figure*}[htpb!]
    \centering
    \includegraphics[width=\linewidth]{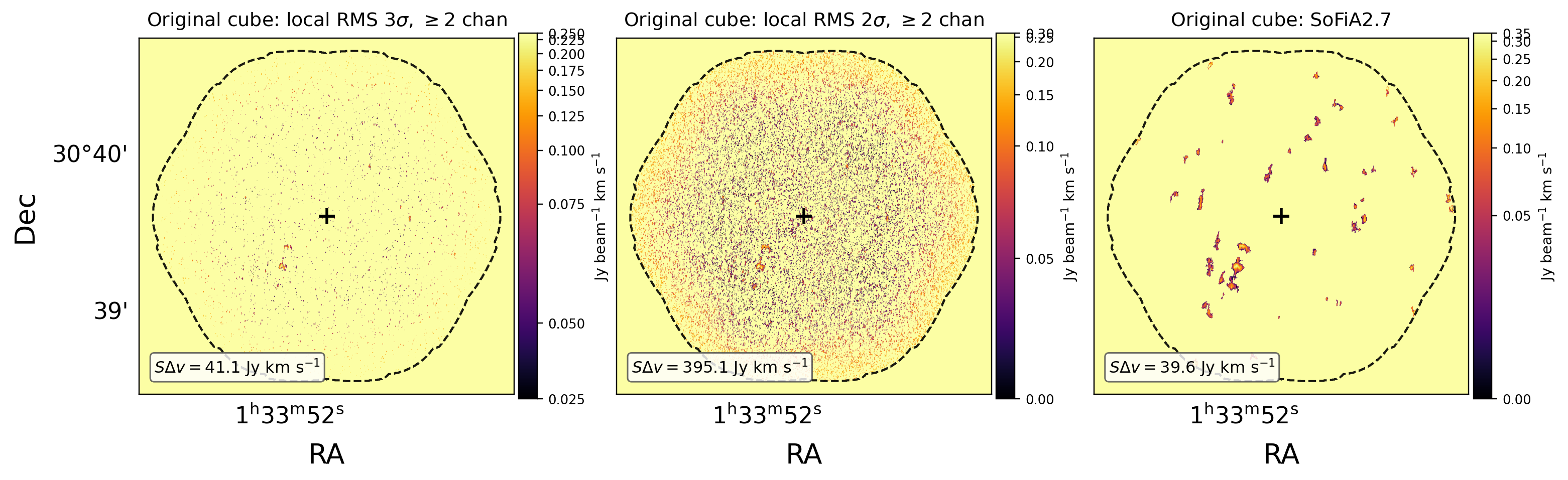}
    \caption{SoFiA integrated-intensity map compared with maps generated
    using local \(3\sigma\) and \(2\sigma\) thresholds. In each
    thresholded map, emission is required in at least two consecutive
    channels.}
    \label{fig:compare}
\end{figure*}

\begin{figure*}[htpb!]
    \centering
    \includegraphics[width=\linewidth]{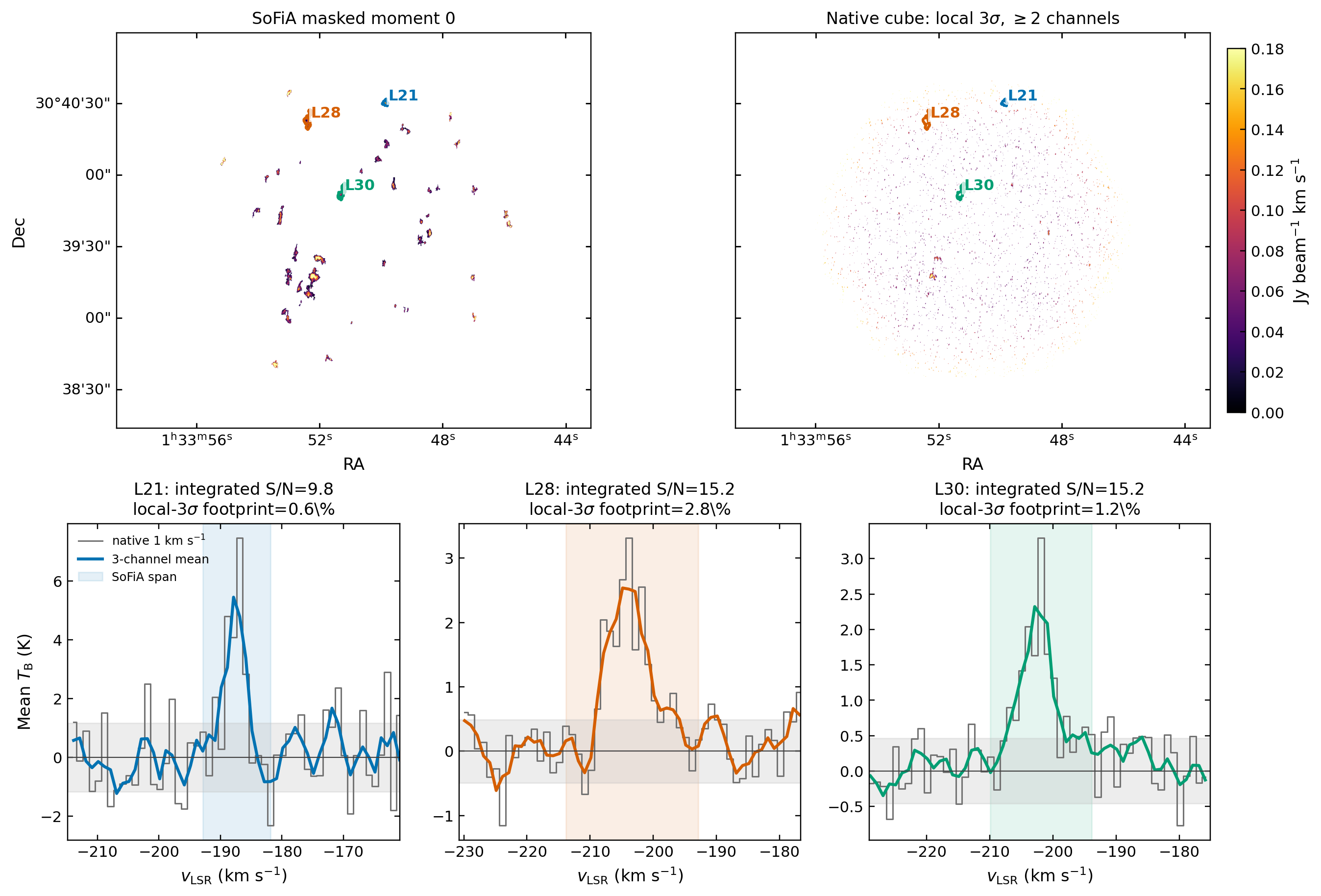}
    \caption{Why a reliable SoFiA source can appear absent from a directly
    clipped moment map. \textit{Top:} SoFiA-masked moment 0 and the native-cube
    local-$3\sigma$ moment map, displayed with a common linear scale. Colored
    outlines mark three representative SoFiA footprints. \textit{Bottom:}
    footprint-averaged native spectra (gray steps) and three-channel running
    means (colored curves). Colored bands show each source's SoFiA velocity
    span and gray horizontal bands show the footprint-averaged rms. The
    integrated S/N is measured by SoFiA over the 3D source mask; the quoted
    local-$3\sigma$ percentage is the fraction of the 2D footprint retained
    by the direct consecutive-channel clip. Spatial and spectral averaging
    reveals coherent lines even where few individual native pixels pass the
    per-channel threshold.}
    \label{fig:sofia_spectra}
\end{figure*}

Figure~\ref{fig:sofia_spectra} illustrates the first effect for three representative detections in the M33 velocity range. In the native-channel local-$3\sigma$ map, only 0.6\%, 2.8\%, and 1.2\% of the L21, L28, and L30 SoFiA footprints, respectively, contain retained pixels. Nevertheless, their footprint-averaged spectra show coherent lines over the velocity intervals selected by SoFiA. The catalog integrated S/N values are 9.8, 15.2, and 15.2, and the footprint-averaged peak-to-rms ratios are 6.4--7.2. These spectra demonstrate that the apparent map ``non-detections'' result from comparing different detection statistics rather than from an absence of line emission.

As a more stringent catalog-level sensitivity check, we also ran the same {\sc pycprops} decomposition on a 1.5\arcsec-smoothed cube using a local $2.5\sigma$ mask with emission required in two consecutive channels. Only five objects pass the finite-deconvolution and quality criteria, compared with 23 for the SoFiA mask. Their velocity dispersions (1.49--2.93 \kms) and surface densities (50--540 $M_\odot\,{\rm pc}^{-2}$) overlap the SoFiA-selected ranges, although their radii are larger (3.5--7.7 pc), as expected after smoothing. The alternate catalog is too small to repeat the SE-versus-field comparison and lacks equivalent bootstrap errors. We therefore retain the SoFiA catalog, explicitly describe it as mask-defined, and treat its population statistics as secondary environmental context.

\section{Spectra of the SE knots and neighboring structures}\label{append:spectra}

Figure~\ref{fig:neighbor_spectra} compares the CO(1--0) spectra of S1--S3 with those of L12 and L9. The spectra are displayed in 2 \kms\ bins for clarity, while the quoted Gaussian fits are obtained from the native 1 \kms\ channels. The S1--S3 spectra are extracted with matched \(3\arcsec\)-radius apertures centered on the three knots. Because L12 and L9 are irregular and do not have equivalent knot centers, their spectra are averaged over their fixed two-dimensional SoFiA footprints. The absolute brightnesses of the two groups should therefore not be compared directly; the figure is intended to compare centroid velocities and line shapes.

Single-Gaussian fits give centroid velocities of \(-168.68\), \(-168.64\), and \(-168.56\) \kms\ for S1, S2, and S3, respectively, with fitted dispersions of 3.08, 2.28, and 2.97 \kms. L12 has a compatible centroid of \(-169.33\) \kms\ but a broader, irregular profile with a fitted dispersion of 5.52 \kms. This profile may include spatial velocity variations across its footprint, and it does not establish a direct physical connection to S3. L9 is centered at \(-167.13\) \kms\ and has a narrower fitted dispersion of 1.44 \kms. Together with the gap in the S3--L12--L9 PVD (Figure~\ref{fig:pvd}), these differences favor treating L9 as a separate projected structure, although an origin involving the same larger-scale flow or local feedback cannot be ruled out.

\renewcommand{\thefigure}{B\arabic{figure}}
\renewcommand{\theHfigure}{B\arabic{figure}}
\setcounter{figure}{0}

\begin{figure*}[htpb!]
    \centering
    \includegraphics[width=\linewidth]{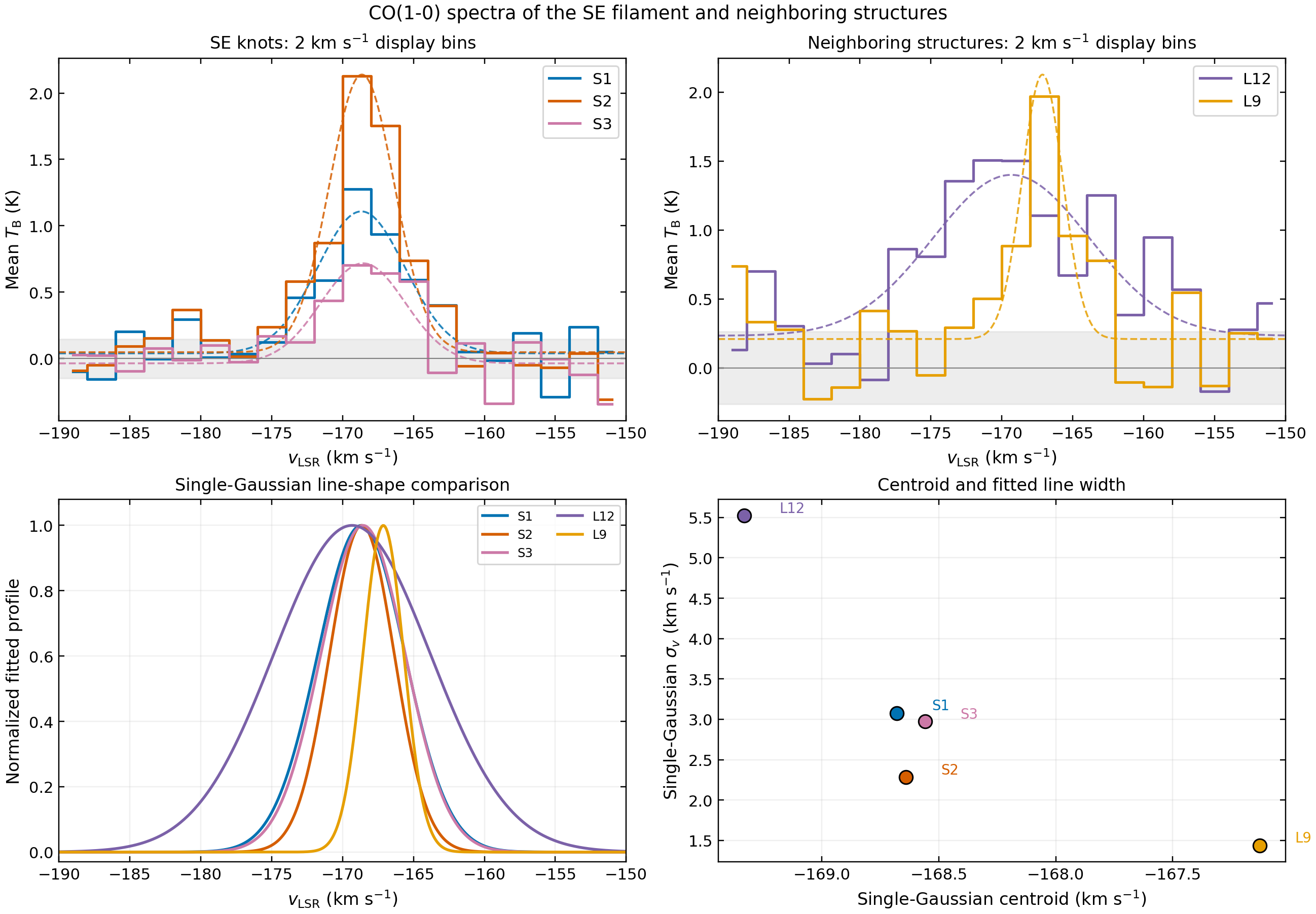}
    \caption{CO(1--0) spectral comparison of the SE filament and
    neighboring compact structures. The stepped spectra are displayed
    in 2 \kms\ bins, the dashed curves show single-Gaussian fits to the
    native 1 \kms\ channels, and the gray bands indicate the
    approximate rebinned rms. \textit{Upper left:} Spectra of S1--S3
    extracted in matched \(3\arcsec\)-radius apertures.
    \textit{Upper right:} Spectra of L12 and L9 averaged over their
    fixed SoFiA footprints. \textit{Lower left:} Peak-normalized
    single-Gaussian models for all five structures. \textit{Lower right:}
   Fitted centroids and velocity dispersions. The similar
    centroids of S1--S3 support their classification as a single clumpy
    filament. L12 has a compatible but broader and less regular
    profile, while L9 is redshifted and narrower. Because the extraction
    regions differ, absolute brightness should not be compared between
    the upper panels.}
    \label{fig:neighbor_spectra}
\end{figure*}

\section{SE-filament channel maps}\label{append:channels}

{Figure~\ref{fig:se_channels} shows CO(1--0) channel maps of the SE region at the original $0.91\arcsec\times0.37\arcsec$ angular resolution. Each panel averages two adjacent 1 \kms\ channels and is made directly from the cube without source masking. Emission from S1--S3 is strongest between about $-173$ and $-163$ \kms\ and shifts systematically across this range, consistent with the transverse structure in the m1 map and PVD.}

\renewcommand{\thefigure}{C\arabic{figure}}
\renewcommand{\theHfigure}{C\arabic{figure}}
\setcounter{figure}{0}

\begin{figure*}[htpb!]
    \centering
    \includegraphics[width=0.88\linewidth]{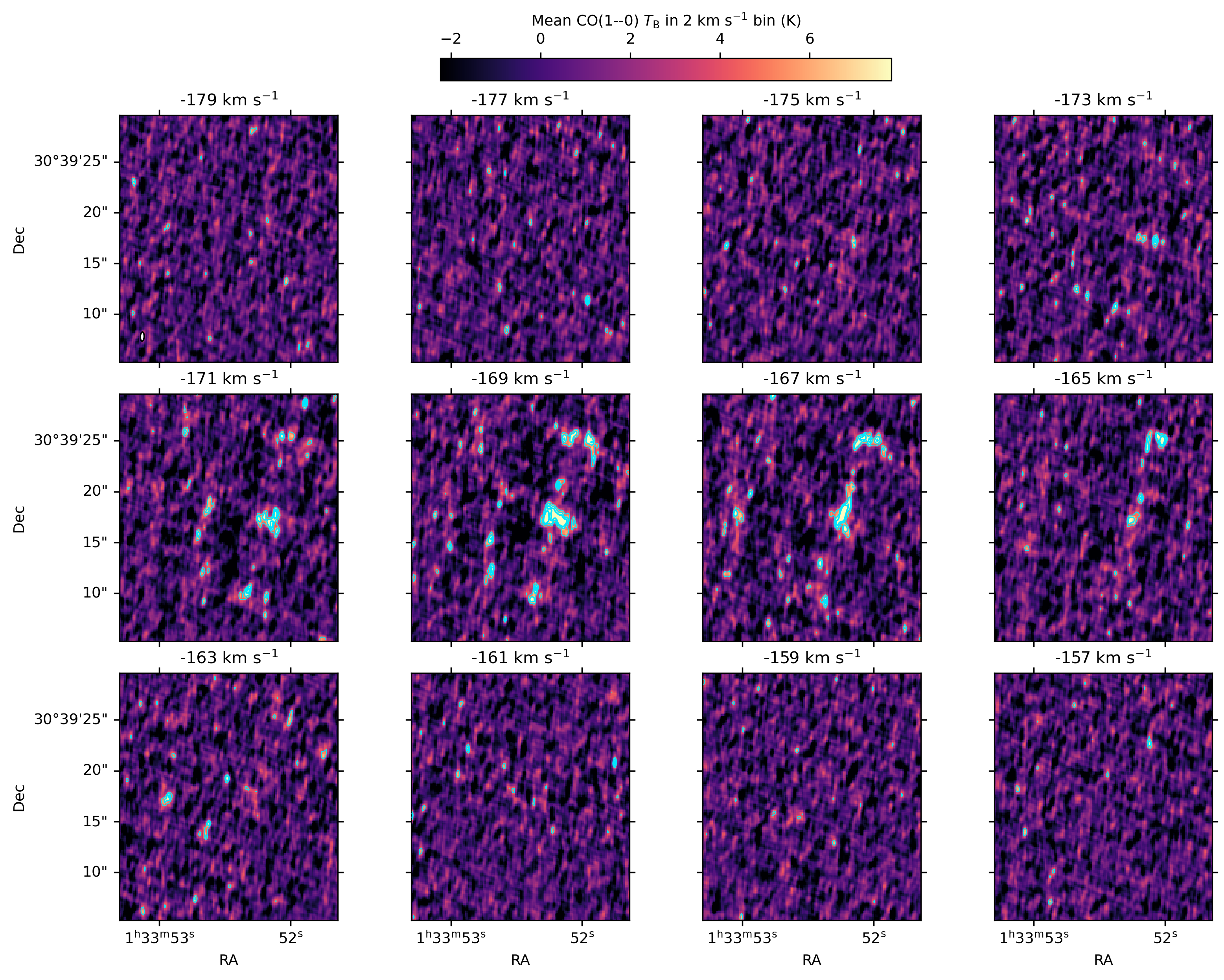}
    \caption{{NOEMA CO(1--0) channel maps of the SE region in twelve 2 \kms-wide bins centered from $-179$ to $-157$ \kms. Colors show the mean brightness temperature, and cyan contours mark 4, 6, and 8 times the binned rms. The synthesized beam is shown in the first panel.}}
    \label{fig:se_channels}
\end{figure*}

\clearpage
\section{Compact-clump catalog}\label{append:catalog}

{Table~\ref{tab:clumps} lists the positions and properties of the 23 compact clumps used in the statistical analysis. The catalog is also provided in machine-readable form.}

\renewcommand{\thetable}{D\arabic{table}}
\renewcommand{\theHtable}{D\arabic{table}}
\setcounter{table}{0}

\startlongtable
\begin{deluxetable*}{lccccccccr}
\tabletypesize{\scriptsize}
\tablecaption{{NOEMA CO(1--0) compact-clump catalog}\label{tab:clumps}}
\tablehead{
\colhead{ID} & \colhead{R.A.} & \colhead{Decl.} & \colhead{$R$} & \colhead{$\sigma_v$} & \colhead{$L_{\rm CO}$} & \colhead{$M_{\rm mol}$} & \colhead{$\Sigma_{\rm mol}$} & \colhead{$\alpha_{\rm vir}$} & \colhead{Region} \\
\colhead{} & \colhead{(deg)} & \colhead{(deg)} & \colhead{(pc)} & \colhead{(\kms)} & \colhead{($10^3$ K \kms\ pc$^2$)} & \colhead{($10^3\,M_\odot$)} & \colhead{($M_\odot$ pc$^{-2}$)} & \colhead{} & \colhead{}
}
\startdata
C026 & 23.465484 & 30.645356 & $3.69\pm0.58$ & $1.39\pm0.20$ & 0.789 & $3.392\pm0.511$ & 79 & 2.19 & Field \\
C047 & 23.457562 & 30.670386 & $2.53\pm0.31$ & $3.07\pm0.29$ & 0.908 & $3.904\pm0.463$ & 195 & 6.39 & Field \\
C065 & 23.463788 & 30.664440 & $3.15\pm0.37$ & $2.72\pm0.29$ & 0.650 & $2.796\pm0.379$ & 90 & 8.76 & Field \\
C077 & 23.468442 & 30.673028 & $3.32\pm0.27$ & $2.15\pm0.32$ & 0.829 & $3.566\pm0.514$ & 103 & 4.50 & Field \\
C089 & 23.458715 & 30.668484 & $2.23\pm0.47$ & $1.81\pm0.42$ & 0.120 & $0.515\pm0.172$ & 33 & 14.80 & Field \\
C090 & 23.458516 & 30.668401 & $1.95\pm0.38$ & $2.36\pm0.34$ & 0.282 & $1.214\pm0.231$ & 101 & 9.35 & Field \\
C094 & 23.475061 & 30.662543 & $2.67\pm0.25$ & $2.79\pm0.31$ & 0.829 & $3.564\pm0.483$ & 159 & 6.10 & Field \\
C095 & 23.457779 & 30.675087 & $2.82\pm0.40$ & $1.80\pm0.41$ & 1.173 & $5.044\pm0.773$ & 202 & 1.89 & Field \\
C112 & 23.451812 & 30.659904 & $2.46\pm0.26$ & $2.92\pm0.25$ & 2.035 & $8.751\pm0.839$ & 460 & 2.50 & Field \\
C114 & 23.452594 & 30.658810 & $2.53\pm0.47$ & $1.92\pm0.38$ & 0.314 & $1.350\pm0.300$ & 67 & 7.25 & Field \\
C117 & 23.453049 & 30.658871 & $2.38\pm0.32$ & $1.91\pm0.41$ & 0.372 & $1.599\pm0.266$ & 90 & 5.67 & Field \\
C124 & 23.467673 & 30.652565 & $4.09\pm0.58$ & $2.83\pm0.61$ & 0.381 & $1.639\pm0.330$ & 31 & 20.83 & SE \\
C126 & 23.468213 & 30.652894 & $3.16\pm0.39$ & $1.60\pm0.26$ & 0.341 & $1.468\pm0.271$ & 47 & 5.79 & SE \\
C127 & 23.467905 & 30.654943 & $1.40\pm0.09$ & $3.27\pm0.25$ & 1.683 & $7.237\pm0.773$ & 1178 & 2.16 & SE \\
C131 & 23.467383 & 30.654790 & $5.38\pm0.20$ & $2.99\pm0.15$ & 5.032 & $21.637\pm1.412$ & 238 & 2.32 & SE \\
C132 & 23.468441 & 30.652777 & $2.44\pm0.23$ & $3.13\pm0.31$ & 0.724 & $3.114\pm0.368$ & 166 & 8.04 & SE \\
C135 & 23.468093 & 30.652711 & $2.62\pm0.33$ & $2.01\pm0.40$ & 0.863 & $3.713\pm0.290$ & 172 & 2.98 & SE \\
C143 & 23.470960 & 30.654900 & $3.59\pm0.46$ & $1.72\pm0.26$ & 0.760 & $3.267\pm0.433$ & 81 & 3.40 & Field \\
C144 & 23.469912 & 30.657570 & $1.70\pm0.50$ & $1.83\pm0.40$ & 0.240 & $1.034\pm0.189$ & 114 & 5.76 & Field \\
C166 & 23.470869 & 30.649681 & $2.76\pm0.25$ & $2.48\pm0.38$ & 0.966 & $4.153\pm0.580$ & 173 & 4.28 & Field \\
C171 & 23.471510 & 30.650710 & $2.74\pm0.30$ & $1.89\pm0.23$ & 0.741 & $3.186\pm0.413$ & 135 & 3.21 & Field \\
C176 & 23.450715 & 30.665050 & $1.95\pm0.37$ & $0.57\pm0.24$ & 0.266 & $1.144\pm0.296$ & 96 & 0.58 & Field \\
C190 & 23.441050 & 30.660963 & $3.18\pm0.42$ & $4.48\pm0.76$ & 2.742 & $11.791\pm1.498$ & 370 & 5.67 & Field \\
\enddata
\tablecomments{{Positions are intensity-weighted centroids. Radii, linewidths, and masses are extrapolated and deconvolved {\sc pycprops} measurements; quoted uncertainties are bootstrap estimates. Masses, surface densities, and virial parameters use \(\alpha_{\rm CO}=4.3~M_\odot~({\rm K~km~s^{-1}~pc^2})^{-1}\). The ``SE'' designation identifies the six clumps associated with the SE filament. A machine-readable version of the table is provided.}}
\end{deluxetable*}

\clearpage
\bibliography{ref_arxiv}{}
\bibliographystyle{aasjournalv7}

\end{document}